# Twistronics in graphene-based van der Waals structures

Ya-Ning Ren[§], Yu Zhang[§], Yi-Wen Liu[§], and Lin He[†]
Center for Advanced Quantum Studies, Department of Physics, Beijing Normal University, Beijing, 100875, People's Republic of China
[§]These authors contributed equally to this work.
[†]Correspondence and requests for materials should be addressed to L.H. (e-mail: helin@bnu.edu.cn).

**The electronic properties of van der Waals (vdW) structures can be substantially modified by the moiré superlattice potential, which strongly depends on the twist angle among the compounds. In twisted bilayer graphene (TBG), two low-energy Van Hove singularities (VHSs) move closer with decreasing twist angles and finally become highly non-dispersive flat bands at the magic angle (~ 1.1º). When the Fermi level lies within the flat bands of the TBG near the magic angle, Coulomb interaction is supposed to exceed the kinetic energy of the electrons, which can drive the system into various strongly correlated phases. Moreover, the strongly correlated states of flat bands are also realized in other graphene-based vdW structures with an interlayer twist. In this article, we mainly review the recent experimental advances on the strongly correlated physics of the magic-angle TBG (MATBG) and the small-angle twisted multilayer graphene. Lastly we will give out a perspective of this field.**

# 1. Introduction

The discovery of graphene has generated an extensive investigation on two-dimensional van der Waals (vdW) materials, among which the bilayer graphene is considered as an ideal model system and has attracted much attention in the past few years. Both experiments and theoretical calculations reveal that the interlayer vdW interactions and band structures of bilayer graphene can be dramatically modulated by the twist between two adjacent graphene layers [1-29]. For the large twist angles of $\theta >$ 5.5°, the two graphene layers are usually electronically decoupled and each layer of them behaves as a monolayer graphene [21, 28-32], except for a small set of angles which yield commensurate structures [2, 16-18, 33]. With decreasing the twist angle, low-energy Van Hove singularities (VHSs) of twisted bilayer graphene (TBG) gradually move closer, accompanied by a pronounced suppression of Fermi velocity due to the strongly interlayer coupling [4-15, 21, 25, 26, 34]. When the twist angle approaches the magic angle (~ 1.1°), the Fermi velocity almost vanishes, resulting in two highly non-dispersive flat moiré bands closely flanking the charge neutrality point [6-11, 25, 35-58]. In the past two years, the TBG near the magic angle has opened a fascinating new chapter in the field of strongly correlated quantum matter [49-85]. An exceptionally wide range of correlated physics, such as Mott insulator, superconductivity, ferromagnetism, and topology, are experimentally observed in the TBG near the magic angle [51, 52, 63-82]. In this article, we first summarize the electronic properties of the TBG within a broad range of the twist angles from 0° to 60°. And then, we mainly focus on the strongly correlated phases in the TBG near the magic angle, following by the comparison of flat-band physics in the small-angle twisted multilayer graphene. After that, the moiré periodic potential induced exotic physics in graphene/hexagonal boron nitride (hBN) heterostructure and the proximity effect induced properties in graphene/transition metal dichalcogenides (TMDs) heterostructures are briefly introduced. At last, we present the outlook and challenges in this field.

## 2. Electronic properties of TBG

Considering that the complex geometry of the TBG can dramatically affect its electronic properties [1-29], we first devote some attention to the atomic structures. Strictly, the periodic moiré superlattice of the TBG only occurs at specific so-called commensurate angles that satisfy [2-7, 14-16]:

$$\theta = \cos^{-1}\left(\frac{3p^2 + 3pq + q^2/2}{3p^2 + 3pq + q^2}\right) \quad (1)$$

where *p* and *q* are coprime positive integers. As shown in Fig. 1, the commensurate superlattice with a long-ranged moiré period in real space is easily formed for the twist angles in the vicinity of 0 ° and 60 °. In contrast, the incommensurate superlattice with rotational symmetry but not a long-ranged translational symmetry is more common for the twist angles around 30 ° [87-96]. Usually, study on the TBG mainly focuses on the commensurate superlattices because the strength of interlayer coupling is expected to introduce many exotic phenomena, which are quite different from that in the components [51-85, 97-115]. Among all, the TBG near the magic angle of approximately 1.1 ° has attracted enormous interest. Many strongly correlated physics, such as Mott insulating state, superconductivity, strange metal (non-Fermi liquid), nematicity, topology, and magnetism for the partially filled moiré flat minibands have all been reported in recent experiments [51-85] (Fig. 1).

With the reduction of the twist angle, the interplay between the vdW interaction energy and the elastic energy of the TBG is expected to result in obviously structural relaxation and atomic reconstruction [115-119]. Specifically, the atomic registry varies continuously across the moiré period between AA, AB, and BA stacking configurations, with the growth of the energy-favorable AB and BA stacking regions and the reduction of the AA stacking region (In the AA stacking region, each top-layer atom locates directly onto a bottom-layer atom. In the AB/BA stacking region, each A/B-sublattice atom of the top layer locates onto a B/A-sublattice atom of the bottom layer, while the B/A-sublattice atoms of the top layer have no partner in the bottom layer) [115-119]. Once the inverse symmetry of the AB/BA stacking is broken by an external displacement field or the substrate-induced staggered potential, it is expected to open a band gap [120-125], accompanied by the different band topology of the AB and BA stacking regions [126-132]. In this case, the topologically protected one-dimensional network of valley Hall states and two pairs of chiral edge modes at the AB-BA domains are expected to appear [97-113, 115], as shown in the bottom left panel of Fig. 1. Moreover, the AB-BA domain and its band structure can be tuned by electrostatic force near the magic angle [105]. The large coherence length of the topological network modes provides a platform to accommodate various interference oscillations. Very recently, the Aharonov-Bohm oscillations under perpendicular magnetic fields are experimentally observed in marginally TBG due to the existence of a triangular network of one-dimensional states [101-103] (top left panel of Fig. 1).

For the large twist angles around 30°, the atomic structure of the TBG is usually incommensurate with only quasi-periodicity symmetry [87-96]. On the one hand, the interlayer interaction of the large-angle TBG was assumed to be negligible due to the lacking of phase coherence, which was thought of as a pair of decoupled monolayer graphene sheets [21, 28-32]. On the other hand, the conventional moiré effective theory, which bases on the interference of the lattice periods [1-19], cannot describe the electronic structures of such an incommensurate stacking. The large-angle TBG has attracted increasing attention only recently when the 30° TBG was successfully synthesized on the 4H-SiC (0001) [88, 133, 134], Pt (111) [89], Cu-Ni(111) [135], and Cu(111) [92, 94, 136] substrates. The 30° TBG, a typical incommensurate configuration with a 12-fold rotational symmetry (Fig. 1), is the first experimental realization of the two-dimensional quasicrystals based on graphene. Contradicting the belief, recent angle-resolved photoelectron spectroscopy (ARPES) [88, 89] and scanning tunneling microscopy (STM) [92] experiments show the signatures of extra mirrored Dirac points inside the Brillouin zone of each graphene layer, together with a gap opening at the zone boundary (Fig. 1). This result demonstrates the strongly interlayer coupling with quasi-periodicity in the 30° TBG. Moreover, the well-defined Landau quantization of massless Dirac fermions is observed in the scanning tunneling spectroscopy (STS) spectra, highlighting the relativistic Dirac fermion quasicrystal of the 30 °TBG [92].

Around the 30 °, there still are a series of discrete large twist angles of the TBG with the commensurate configuration [2, 16-18, 33], as shown in Fig. 1. The angular-dependence conductivity measurement also highlights the commensurate crystalline superstructure of the TBG at the twist angles of 21.8 ° and 38.2 ° [33], arising from the coherent two-dimensional electronic interface state [2, 16]. Theoretically, the TBG with large commensurate angles (such as $\theta = 21.8°$) are predicted to be a generic higher-order topological insulator, which is characterized by the topologically protected corner states [114]. Moreover, for the TBG with a twist angle that is sufficiently close to the commensurate 38.2 °, the alternative commensurate configurations, *i.e.* the gapped exchange symmetry even (SEE) and the gapless exchange symmetry odd (SEO) [2], geometrically frustrate the network of topologically protected counterpropagating chiral modes [137]. The emergent geometric frustration of the system gives rise to flat bands and a Kagome lattice [137], which may be an exciting platform for exploring the spin-liquid physics and other exotic states [138-140]

As we summarized in Fig. 1, the TBG exhibits a variety of exotic electronic properties, which strongly depend on the interlayer twist angle. In the following part, we mainly focus on strongly correlated physics in the magic-angle TBG (MATBG).

## 3. MATBG

### 3.1 Symmetry breaking in the MATBG

In the TBG, a finite interlayer coupling between the two graphene layers leads to the emergence of two saddle points appearing at the intersections of the two Dirac cones, which consequently generate two low-energy VHSs in the density of states (DOS) (inset of Fig. 2). Theoretically, the energy separation of the two VHSs can be roughly estimated as [4, 21-26, 28]:

$$\Delta E_{VHS} \approx \hbar v_F \Delta K - 2t_\theta \quad (2)$$

where $\Delta K = 2|K|\sin(\theta/2)$ is the relative shift of the Dirac points of the adjacent graphene bilayers with the reciprocal-lattice vector $K$ in reciprocal space, $t_\theta$ is the interlayer hopping, $\hbar$ is the reduced Planck constant, and $v_F$ is the Fermi velocity. According to the Eq. (2), the two VHSs move closer to each other with decreasing the twist angles. Here we summarize the energy separations of the two VHSs $\Delta E_{VHS}$ vs $\theta$ of the TBG acquired from the STS measurements [20, 22-28, 55-60, 86], as shown in Fig. 2. Consistent with the Eq. (2), the $\Delta E_{VHS}$ generally decreases with decreasing the twist angles. However, there are a wide distribution of the experimental data, which may result from the variation of the $t_\theta$ in different TBG samples.

When the twist angle is close to the magic angle (~ 1.1°), the quantum interference of electrons in the superlattice potential gives rise to extremely narrow low-energy bands, i.e., the flat bands, and isolates them from higher-energy dispersive bands [6-11, 35-48], as shown in Fig. 3(a). There are three important parameters in the single-particle band structure of the TBG near the magic angle: (i) the energy separations of the two VHSs or the two flat bands (labeled as $\Delta E_{VHS}$), (ii) the bandwidth of each VHS or each flat band, and (iii) the band gap between the flat bands and the upper dispersive bands (labeled as $\Delta_{gap}$). These parameters are systematically measured by different groups in the past few years [20-25, 50-69, 115]. In Fig. 3(b)-(e), we summarized the $\Delta E_{VHS}$, the full-width at half-maximum (FWHM) of the VHSs, and the $\Delta_{gap}$ acquired from STS and transport measurements in the TBG near the magic angle, respectively. Obviously, the measured data vary from sample to sample and exhibit large differences, sometimes even by an order of magnitude in the TBG with almost the same twist angle. This result highlights the complexity and lack of reproducibility even in the single-particle band structure of the TBG near the magic angle..

When the Fermi level lies within the flat bands in the MATBG, the Coulomb interactions between electrons, which can be estimated by $U \sim e^2/\epsilon l_m$ ($\epsilon$ is the effective dielectric constant, and $l_m$ is the moiré period), become important with respect to the kinetic energy set by the bandwidth $W$ of the flat bands [143-145, 187-191]. Therefore, the system is expected to exhibit various strongly correlated phases. Indeed, recent STS measurements have shown obvious signatures of many-body correlations in the TBG near the magic angle [52-60, 62, 84, 85]. When the two flat bands in the MATBG are fully occupied or empty, the STS spectra show sharp peaks associated with the two nearly flat conduction and valence bands, which are consistent with the single-particle pictures [52-58]. When one of the flat bands is partially occupied (not just at commensurate fillings), the spectra show pronounced bands splitting together with the broadening of both the two flat bands [52-58], indicating the importance of interaction-induced symmetry breaking states.

One of the major differences in the STS spectra extracted from different research groups is worth highlighting here. For the devices reported by Kerelsky (1.15°) *et al.* [57] Xie (1.01°) *et al.* [55] Choi (1.04°) *et al.* [56] and Jiang (1.07°) *et al.* [58], the partially occupied flat bands split into two peaks with their energies locating on either side of the Fermi level in the spectra (as schematically shown in Fig. 4(a)). Other groups, such as Ren *et al.* [59], reported that the partially filled flat band spontaneously splits into four peaks in a 1.49° TBG, as shown in Fig. 4(b). Because of the fourfold spin–valley flavour degeneracies of each flat band, the strong electron–electron interactions are supposed to drive the degenerate bands splitting into sub-bands in the framework of Hartree–Fock approximation [143, 176, 186]. The behaviours of the spontaneous symmetry-breaking states of electrons in the MATBG are analogous to the quantum Hall ferromagnets in Landau levels (LLs) of graphene monolayer [201, 202], in which the isospin SU(4) symmetry of each LL is lifted due to the strong electron–electron interactions [194-200]. And simultaneously, the partially occupied LL splits into subset at all the integer fillings [201-210]. Therefore, the splitting flat bands in the MATBG can be naturally attributed to the correlation-induced valley and/or spin polarized states, suggesting the abundant many-body ground states in the MATBG.

**3.2 Correlated insulator and superconductivity in the MATBG**

A central question in the TBG near the magic angle is the nature of the observed many-body ground states, which have been widely studied in the past two years [51-85]. To date, many groups have examined the longitudinal resistance (conductance) in the

TBG near the magic angle as a function of carrier densities, temperatures, magnetic fields, and pressure via the transport measurements [63-81], and a zoo of strongly correlated phases, such as correlated insulating states and superconductivity, are clearly captured [63-81]. Similarly, the doping-dependent STS intensities at the Fermi level can also mimics the conductance measured with the transport where only charge carriers near the Fermi level have contributions [53-56, 58]. Figure 5 summarized representative results obtained in transport measurements [68, 70-72, 78] and STM measurements [54-56, 58] in the TBG near the magic angle. A common signature of insulating states appears at most or all integer occupancies of the fourfold degenerate flat conduction and valence bands, i.e., at moiré band filling factors ν, where ν = 0, ±1, ±2, ±3 indicate the number of charge carriers per moiré superlattice unit cell. Close to the certain insulating states, superconducting domes are observed below critical temperatures via the transport measurements, as illustrated schematically in Fig. 5 [68, 70-72, 78]. Importantly, there are significant differences in the acquired phase diagrams in different experiments, as we will discuss subsequently.

First, some groups show the existence of correlated insulating states at all integer moiré band fillings [72] while other groups find semimetallic or metallic behavior at the same integer moiré band fillings [63-71, 73-81]. The reasons for this difference might be attributed to the unavoidable strains [55-57, 104, 211], inhomogeneous twist angles [77, 79, 82, 212], the aligned or unaligned devices of the MATBG with respect to the underlying substrate [177-183, 213-215], and the thickness of a dielectric spacer layer (typically hBN) [70, 71, 75, 216]. On the one hand, the electronic properties of the MATBG is very sensitive to the lattice relaxation and heterostrain [115, 117, 211], and the strain values of the devices during the sample preparation processes are indeed different according to several STM experiments [52-60, 62, 84, 85]. On the other hand, the device setups employed by different groups show great differences. For instance, Cao *et al.* encapsulated the MATBG by hBN flakes with the thicknesses of about 10-30 nm [65, 68]. In contrast, the thicknesses of hBN in the fabricated devices are 30-50 nm, 25-50 nm, and 10 nm for the experiments in Saito *et al* [70], Polshyn *et al* [63], and Lu *et al* [72], respectively. Recent experiments show that the dielectric environment of the MATBG, i.e. thickness of a dielectric hBN layer, has a strong influence on the screened electron-electron interactions, which in turn have significant influence on the obtained phase diagram [70, 71, 216]. In addition, a recent superconducting quantum interference device fabricated on the apex of a sharp pipette (SQUID-on-tip) measurement also highlights the unavoidable varying degree of twist-angle homogeneity in the TBG samples [77]. The large variations in the obtained correlation phase diagrams in the TBG near the magic angle are reasonable and expected with considering the lack of reproducibility even in the single-particle band structure in this system.

Second, superconducting signatures are not always found at the same doping levels in the TBG with almost the same twist angle in the transport experiments [63-81] and no signature of superconductivity is observed in the STM measurements [52-58] up to now (Fig. 5(b)). In the transport experiments, the superconductivity is usually deduced from a (near) zero-resistance state that is rapidly suppressed with increasing temperatures and magnetic fields, accompanied with a field-dependent critical current that is consistent with Berezinskii–Kosterlitz–Thouless (BKT) theory [68, 69, 71, 72, 76]. At present, the superconductivity is lack of reproducibility in the transport measurement and only a few samples exhibit the zero-resistance state. It is even more abnormal that all the STS spectra show no signature of superconducting gap in the MATBG up till now [52-58]. The local superconducting signals should be much more accessible in the STM studies since the spatial resolved STM and STS measurements can efficiently avoid the impurities and inhomogeneous of the sample [217-220]. This reminds us to treat the (near) zero-resistance phenomena in transport measurements more critically, since other mechanisms in the transport measurements such as ballistic transport may also lead to similar zero-resistance phenomena [221]. In addition, recent studies pointed out that the similar sudden drop in resistivity and nonlinear $I$–$V$ characteristic were also observed in twisted double-bilayer graphene, and these phenomena may possibility originate from the spin-polarized correlated state [222]. Further measurements should be carefully carried out in the near future to rule out any other possibilities.

On the other hand, if the (near) zero-resistance feature indeed comes from the superconductivity, the origin and mechanism of the superconductivity in the MATBG is another key issue for researchers to address. So far, there are mainly two different opinions on the relationship between superconductivity and insulating behavior in the MATBG. One is that the superconductivity and insulating behavior are connecting [66-68, 72-78], the other is that they are competing [63, 69-71]. For the research groups such as Cao *et al*. [67, 68, 76], their experimental data show that both of the superconductivity and insulating state arise from strongly electron-electron interactions and are closely connecting in the MATBG. Their experiments indicate a tantalizing resemblance to that of cuprate superconductors for the following three reasons: (1) The phase diagrams show that the superconducting domes are always in close proximity to the insulating states [68, 152, 171, 223-225]; (2) A broad regime of temperature-linear resistivity is observed for a range of fillings near the correlated insulator, and the ratio of critical temperature to the Fermi energy is relatively large [67, 226-231]; (3) A strongly anisotropic phase appears in a wedge above the under-doped region of the superconducting dome, and the superconductivity exhibits an anisotropic response to the direction of in-plane magnetic fields, which reveal a nematic pairing state across the entire superconducting dome [76, 173, 185, 232-241]. However, for the results obtained by the research groups such as Saito *et al*. [70] and Stepanov *et al*. [71], the superconductivity and correlated insulator are competing with each other in the MATBG: the insulating phase arises from electron-electron interactions while superconductivity arises from conventional phonon-mediated pairing [165-170]. In their experiments, the correlated insulating states at $\nu = \pm 2$ are substituted to metallic states by decreasing the thickness of dielectric environment to independently suppress the Coulomb energy [70, 71]. However, the superconductivity still persists in the absence of the correlated insulators and even takes over the phase space vacated by the correlated insulators [69-71]. Moreover, the superconductivity can survive to a larger detuning of the twist angles [69-71]. In this framework, the electric pairing between weakly interacting band electrons, since any tendency of the electrons to form an insulating phase would suppress superconductivity because fewer electrons would be available to pair. In addition, conventional phonon-mediated pairing in graphene-based systems with a high DOS at the Fermi level is a realistic possibility, such as in an intercalated graphite [242-246]. Despite much theoretical work [40-42, 141-173,

[187], the microscopic mechanism of the pairing interaction and the symmetry of the superconducting order parameter still have not been reached consensus, which needs to be continuously explored.

**3.3 Nematicity in the MATBG**

In the single-particle theoretical picture, the moiré periodic potential in the TBG leads to the spatial variation of the DOS, which are expected to exhibit the same or inverted contrast as atomic topographies in the STM images [1-3, 7-9]. The Coulomb interactions can efficiently break the symmetry of electronic wavefunctions [173, 185, 239-241], resulting in the spatial redistribution of the DOS within each moiré superlattice. Experimentally, the spatial-resolved DOS can be directly visualized via the conductance maps in the STM studies [53, 56-60, 192]. Figure 6(a) shows a schematical atomic structure of two misoriented graphene sheets with the twist angle θ, which exhibits the $C_6$ rotational symmetry. When the twist angle is far from the magic angle, the conductance maps at energies around the VHS peaks show the same features as topographies and maintain the $C_6$ symmetry [58, 60, 192]. Similar $C_6$-symmetry conductance maps are also observed in the MATBG when the two flat bands are fully filled or empty (Fig. 6(b)) [58]. However, when the Fermi level locates at the charge neutrality point (CNP) of the MATBG, the low-energy conductance maps show an obvious correlation-induced charge redistribution in each moiré, which suppresses the DOS at AA regions and highlights the domain wall areas [56]. Particularly, the conductance maps at the energies of two flat bands show pronounced anisotropy, reducing the symmetry from the initial $C_6$ to $C_2$ with the symmetry axes perpendicular to each other [53, 56, 60, 175].

When one of the flat bands (VHS peaks) in the MATBG is partially filled, the wavefunctions for both the flat bands change dramatically, resulting in different symmetry breaking phases for different moiré band fillings [51-59, 62-84, 182-188]. For the partially filled flat band, the DOS for ν = 1 are no longer localized at the AA regions. In contrast, the DOS for ν = 2 are still localized at AA regions, accompanied by the broken of $C_6$ rotational symmetry [53] (as schematically shown in Fig. 6(c)). The stark differences between ν = 1 and 2 fillings indicate that they are intrinsically different correlated phases [53]. These results are well consistent with the fact that the ground state of ν = 1 is both spin and valley polarized, while the ground state of ν = 2 is either spin or valley polarized [176-178, 180-183]. Moreover, the charge redistributions in partially filled flat band prefer to result in the stripe charge order [53, 57-59], suggesting a high susceptibility towards nematicity [173, 185, 239-241]. This result is similar to the cuprate superconductors, where the superconductivity is intimately connected to the nematic phase and pseudo-gap state [232-237]. However, we should still be much careful about this viewpoint. On the one hand, only a small amount of strain, which is almost unavoidable in the TBG, can introduce additional broken of symmetries [211, 239, 247]. Although the strain-induced trivial symmetry-breaking state is supposed to be uniform with different energies and fillings, there still exists the possibility that the coexistence of both strain and unaccounted electronic correlations stabilize the system as the observed ground state [175, 248]. On the other hand, the domains of stripe charge order with different orientations are expected to appear in the nematic phase, which is still experimentally lacking in the MATBG up till now. In addition, theoretical calculations have predicted abundant competing quantum states and correlated ground phases in the MATBG, which strongly depend on the subtle atomic structures and sample preparation processes [170-191]. Therefore, the intrinsic ground states and fundamental effects in the MATBG need to be further carefully studied.

**3.4 Topology and Magnetism in the MATBG**

Recent experiments on ν = 3 filling of the conduction band in the TBG near the magic angle reveal quite distinct correlation-induced quantum states, such as metallic phase [69-71], thermally activated trivial insulator [63, 72-74, 76-78], and (quantized) anomalous Hall effect [64, 81]. These differences highlight the extreme sensitivity of symmetry-breaking phases to the fine atomic structures of the TBG and the surrounding atmosphere. Among all, the (quantized) anomalous Hall effect, which is supposed to be energy-favored when the MATBG is closely aligned to the hBN layers [64, 81, 178-183, 213-215], is of great interest and attracts much attention. Sharpe *et al*. observed a giant anomalous Hall (AH) effect at around ν = 3 of the TBG near the magic angle (1.17º) that is on top of hBN layers with the corresponding angle of 0.81º, and the Hall resistance $R_{xy}$ displayed hysteresis loops under external magnetic fields [81]. Moreover, they acquired the chiral edge currents via the nonlocal transport measurements, which are the characteristic of topological phases [81]. Very recently, Serlin *et al*. not only well repeated the hysteresis loops, but also realized the switch of the magnetization via a small charge current [64]. More importantly, they observed a well-quantized resistance $R_{xy} = h/e^2$ under zero magnetic field at ν = 3 of the hBN-aligned TBG near the magic angle (1.15º), demonstrating the realization of a quantized anomalous Hall (QAH) state in this system [64] (Fig. 7).

The observation of hysteresis loops at ν = 3 filling of the hBN-aligned TBG near the magic angle demonstrates a clear experimental evidence of the ferromagnetic state [64, 81, 82]. Due to the extremely weak spin-orbit coupling and the absence of magnetic atoms in graphene system, the ferromagnetism in the TBG near the magic angle is attributed to the purely orbital magnetism, with an easy-axis anisotropy arising from the graphene bands [177-183, 213, 215]. Recent STM and superconducting quantum interference device (SQUID) measurements also demonstrate the existence of a large orbital magnetic moment of several $\mu_B$ per electron per moiré unit cell in the MATBG [62, 82, 85]. Moreover, the coexistence of (quantized) anomalous Hall effects and ferromagnetic insulating states is reminiscent of the Chern insulator where the longitudinal resistivity $R_{xx}$ approaches to zero and the Hall resistivity is quantized to $R_{xy} = h/Ce^2$ (*C* is the Chern number arising from the Berry curvature of the filled bands) [64, 81, 179-181]. Theoretically, the achievement of a Chern insulator should open up a topologically nontrivial gap, *i.e.* the moiré miniband carries a nonzero valley-contrast Chern number [179-181, 214]. In the TBG near the magic angle, the atomic relaxation leads to the low-energy flat band isolating from the higher-energy dispersive bands by an energy gap [45-48, 115, 117]. Although the combination of $C_{2z}$ symmetry (a twofold rotational symmetry around the out-of-plane axis) and time-reversal symmetry in pristine TBG interchanges the two valleys and forbids the Berry curvature of the low-energy flat bands, the underlying near-aligned hBN can open up a band gap at the Dirac point of graphene and efficiently break the $C_{2z}$ symmetry [64, 81, 178-183, 213-215]. Moreover, for ν = 3 filling flat band in TBG near the magic angle, the strongly *e-e* interactions are expected to spontaneously break both the spin and valley symmetries, resulting in the electrons polarizing into a spin- and valley-resolved moiré miniband [176-178, 180-182]. Therefore, a valley Chern number of C = 1 is generated, resulting in the energy-favored QAH state at ν = 3 filling flat band of the hBN-aligned MATBG [64]. Very recently, several groups also observed the field-driven phase transition of valley-projected bands from zero Chern numbers to nonzero Chern numbers at ν = ±1 and -2 fillings of the hBN-non-aligned MATBG [51, 52, 71-74, 80], highlighting the possible competition between trivial insulator and Chern insulator. However, the primary cause for the diversities still remains to be determined.

The discovery of the QAH state in the MATBG provides a new insight to search for high-quality QAH materials. In traditional QAH materials such as $(Bi,Sb)_2Te_3$, the broken of time-reversal symmetry and the realization of topologically nontrivial Chern bands are attributed to the external magnetic dopants [249-254]. Hence, the critical temperature of QAH is much lower than the magnetic ordering temperature, because it is strongly limited by the inhomogeneous structures [249-254]. In contrast, the intrinsic aligned MATBG-hBN processes the necessary ingredients for the QAH state, offering a less disordered system to realize a more robust quantization [64].

**4. Multilayer graphene**

As we shown above, the ratio between Coulomb interactions and kinetic energy (bandwidth) is a key parameter to determine the strength of electronic correlations [143-145, 187-191]. One of the new perspectives for extending the correlated behaviors is to seek for the graphene-based superlattices with a tunable bandwidth. Twisted double-bilayer graphene (TDBG), which consists of two sheets of untwisted Bernal-stacked bilayer graphene stacked together with a twist angle θ, is an idea candidate for investigating the novel correlated physics. On the one hand, the Bernal-stacked bilayer graphene shows parabolic band dispersion [255-257], and the band gap opens up at the charge neutral point with the increasing perpendicular electrical displacement field [120-125], resulting in the TDBG inheriting this electrically tunable flat bands [222, 258-268]. On the other hand, the flat bands in TDBG emerge over a wide range of twist angles and are much narrow owing to the gap [222, 258-269]. This distinguishes from the TBG that the flat bands only emerge in the vicinity of the magic angles [6-11, 25, 35-58]. Therefore, the TDBG is supposed to exhibit a much richer phase diagrams that are highly sensitive to both the twist angle and the application of an electric displacement field (Fig. 8).

Recent transport experiments highlighted the correlated insulating behaviour for TDBG with the twist angle $\theta \lesssim 1.3°$ at $\nu$ = 1, 2, 3 fillings of the moiré unit cell. The correlated insulating behaviour can be sensitively switched on and off by the displacement field [222, 258-262]. Among all, the $\nu$ = 2 correlated insulating state is particular. In stark contrast to the MATBG, the $\nu$ = 2 insulating gap of the TDBG shows a linear response to the parallel magnetic fields and yields the g-factor that is close to $g$ = 2, suggesting the signature of the spin-polarized ground state [222, 259, 260] (Fig. 8). With the twist angle decreasing to 0.84°, multiple correlated states appear at the half filling of each high-energy flat bands, all of which are tunable by the displacement field as well [259]. Similar to the MATBG, Liu *et al*. [222], Cao et al. [259] and Shen et al. [260] observe half- and quarter-filling correlated states in the TDBG system. However, the spin-polarized ground states at half-filling, the drops in resistivity of metallic states at low temperature and their nonlinear I-V characteristics indicate that it exhibits more complex and richer correlated phases than the MATBG system [222, 259, 260]. Untill now, not many explicit zero-resistance experiments have been presented or repeated. Liu *et al*. conjecture that the critical transition is a result of Cooper-pair formation, but the possible superconductivity is developed only in the 1.26° device [222], and Lee *et al.* argue that it is a spin-triplet superconductivity from pairing between opposite valleys [265]. However, because of the absence of well-defined critical current, no saturation to a normal state and the nonlinear I-V outside the halo, He *et al.* conjecture that the larger nonlinear I-V may be formed by Joule heating, and the reversals in the sign of the Hall coefficient indicates that spontaneous symmetry breaking is the origin of resistivity drops [261]. Therefore, more experiments should be taken to reveal the underlying mechanism in the TDBG system. Moreover, when the twist angle is only about 0.1°, it is energetically favorable for commensurate regions to maximize in area, leading to a significant structural reconstruction of the triangular domains that alternate with ABAB- and ABCA-stacked regions [270-272]. This result is similar to the tiny-angle TBG [115-119]. It's worth noting that the ABCA-stacked regions are a good region to study the correlated physics of itself due to the band structure $E(k) \propto k^4$ [273-277].

ABC-stacked trilayer graphene (ABC-TLG) is also a great candidate for achieving the flat-band physics due to the low-energy cubic dispersion relation of $E(k) \propto k^3$ at the K point in the first Brillouin [278-282]. When placing the ABC-TLG onto the hexagonal boron nitride (hBN) with a large moiré superlattice, the moiré periodic potential can lead to the extremely narrow electronic minibands, which are expected to exhibit abundant strongly correlated physics [56, 178, 283-290]. Recent transport experiments show the signatures of the Mott insulating states at 1/4 and 1/2 fillings of the hole minibands in the ABC-TLG/hBN heterostructures at a large vertical displacement field |D| [284]. In the slightly electron and hole-doped 1/4-filling Mott insulating state, the superconducting domes emerge at the temperature below 1 kelvin [56]. Moreover, the topology of the ABC-TLG/hBN moiré miniband shows greatly displacement-field-dependent. For the displacement field D > 0, the hole miniband is topological with a nonzero Chern number of C = 2 at the 1/4 filling. In contrast, reversing the displacement field to D < 0 can lead to a topological trivial moiré miniband (C = 0) [285]. These experimental results have been followed by a number of theoretical studies on the exotic properties of the flat bands in the ABC-TLG/hBN heterostructures [178, 286-290]. However, the possible mechanism of the superconductivity is still full of debate up till now. In addition to the intrinsic flat band of this system, the magnetic field can condense the energy band into more flatten LLs, and the interactions of electron can be revealed by the broken-symmetry states of LLs, which indicates that it is an ideal system to study the novel correlated phases in the quantum Hall regime [278].

More recently, the twisted TLG is also considered as a perfect system to study the emergent electronic correlations of the engineered flat bands[291-303]. In the twisted monolayer graphene-bilayer graphene (tMBG), i.e. the monolayer graphene rotates with respect to the Bernal-stacked bilayer graphene, experimental results show the correlated insulating states at fillings ν = 1, 2, 3 electrons per moiré unit cell when 1.2 ≲ θ ≲ 1.4° [292-294]. The correlated insulating states in the tMBG can be switched on and off by a displacement field D with an asymmetric response to D, highlighting the electron-hole asymmetry in this system [292-294]. In the vicinity of ν = 1 correlated insulating state, there is the signature of superconductivity [294]. Moreover, the quantized anomalous Hall effect with $R_{xy} = h/2e^2$ is observed at ν = 1 and 3 fillings, suggesting the spontaneous valley-polarized bands with the Chern number of C = 2 [293]. These results have been theoretically verified recently [295-298]. In addition, abundant strongly correlated physics such as superconducting, topological, magnetic, insulating, and metallic states and their interactions are also widely predicted in the twisted TLG where one of the three layers is twisted by a small angle relative to perfect AAA, ABA, or ABC stacking [298, 299] and even in twisted multilayer graphene [300-306], which need further research.

## 5. Graphene with other 2D materials

The lattice mismatch between graphene and the underlying substrates gives rise to the moiré periodic potential, which can significantly modify the electronic features of pristine graphene. Especially, the graphene/hBN heterostructure has attracted much interest. On the one hand, hBN can be used as an ideal dielectric substrate, which not only couples weakly to the graphene, but also significantly suppresses the charge inhomogeneities and greatly improves the mobility of charge carriers in graphene [307-311]. Therefore, the graphene/hBN heterostructure enables the experimental probing of rich intrinsic physics in graphene such as the relativistic quantum Hall effect (QHE), fractional QHE, and more exotic correlated states [207, 312-318]. On the other hand, hBN can act as an active component of the vdW heterostructures and realize a moiré superlattice structure together with graphene due to the existence of lattice mismatch and related rotation between them [319]. In this case, the electronic properties of graphene can be strongly modified and the new superlattice Dirac points arise, which may be accompanied by the energy gap opening at the charge neutrality [320-322]. Moreover, the long-wavelength graphene/hBN moiré pattern provides a unique platform to experimentally realize the stunning self-similar recursive energy spectrum, i.e. Hofstadter butterfly [323-325]. The plasmon and phonon polaritons with exceptional properties are also expected to be achieved in these regimes [326, 327]. Moreover, the twisted hBN/graphene/hBN heterostructure can host a second-order moiré superlattice with an even larger periodicity, which may result in a highly reconstructed graphene band structure featuring multiple secondary Dirac points [328]. Recent experiments report the tunable on/off bandgap at the original and secondary Dirac points by rotating the partial component of the heterostructure, suggesting a tunable transition between the absence or presence of inversion symmetry in the twisted hBN/graphene/hBN heterostructure [329]. This result provides an insight for engineering the band structure of the vdW system.

Recent advances in fabrication techniques have made it possible to realize the vdW heterostructures with the components of graphene and TMDs materials [193, 330, 331]. The (twisted) vdW heterostructures offer a unique platform for addressing many fundamental physics that the individual constituent layers may not have [332-334]. One significant feature that all TMDs have in common is the strong intrinsic spin-orbit coupling (SOC), with the order of 100 meV in the valence band and 10 meV in the conduction band [335-340]. As for the graphene/TMDs heterostructures, the proximity-enhanced SOC in graphene with several meV is expected to achieve active graphene spintronics such as the spin Hall effect and the inverse spin galvanic effect even at the room temperature [341-350]. Moreover, the large proximate SOC can efficiently generate the band inversion of bilayer graphene (BLG) in the BLG/TMDs heterostructures [350], and can also result in the gate-tunable transitions between topological and trivial insulating states in the multilayer graphene (MLG)/TMDs heterostructures [351]. Another distinctive electronic property of typical TMDs such as $NbSe_2$, $NbS_2$, $TaSe_2$, and TaS$_2$ is the superconductivity at low temperature [352-355]. The strong SOC induced spin splittings cause the superconducting pairing to be of the Ising type, which is extremely robust to external in-plane magnetic fields [352, 356, 357]. Previous *ab* initio calculations show a large superconducting pairing can be induced into the graphene layer despite the large mismatch between the lattice constants of graphene and NbSe$_2$, and the proximity-induced superconductivity in graphene can still maintain within a wide range of twist angles due to the large size of the Fermi pockets in NbSe$_2$ [358]. Experimentally, the proximity-induced superconducting gap in BLG of the BLG / NbSe$_2$ heterostructure and the Andreev reflection at their junction are clearly observed [359]. Moreover, recent STM experiment captures the periodically modulated pseudo-magnetic field in graphene/ NbSe$_2$ moiré superlattices, which reminds us a new way to design the flat electronic bands in graphene [360]. In the future, the proximity-induced various electronic properties should be carefully studied relying on available degrees of freedom such as the twist angle and strength of interlayer coupling.

**6. Conclusions and perspectives**

In summary, the TBG is expected to exhibit an exceptionally wide range of physical phenomena, which dramatically depends on the twist angles. Especially in the TBG near the magic angle, many outstanding issues in condensed matter physics such as Mott insulating state, superconductivity, strange metal, nematicity, topology, and magnetism are totally achieved for the partially filled moiré flat minibands. Moreover, much richer strongly correlated phase diagram is realized in twisted multilayer graphene due to the additional tunable degree of freedom of the electrical displacement field. However, there still exist many fundamental mysteries hidden in the graphene-based twisted materials. One of the key questions is whether there is any relation between the zero-resistance feature in transport measurement and superconductivity. If yes, what is the origin and mechanism of the superconductivity? Moreover, the ground states of these materials at different fillings of the moiré flat band and their phase transitions under external magnetic fields are key questions for researchers to address.

Corresponding theoretical calculations need to be carried out to capture the intrinsic physics and predict new phases in this regime. In addition, the results in the MATGB can also extend to other flat-band system, such as the strained graphene with flat bands.

**References**

[1] Landgraf W, Shallcross S, Türschmann K, Weckbecker D and Pankratov O 2013 *Phys. Rev. B* **87** 075433

[2] Mele E J 2010 *Phys. Rev. B* **81** 161405

[3] Shallcross S, Sharma S and Pankratov O 2013 *Phys. Rev. B* **87** 245403

[4] Lopes dos Santos J M B, Peres N M R and Castro Neto A H 2007 *Phys. Rev. Lett.* **99** 256802

[5] Shallcross S, Sharma S, Kandelaki E and Pankratov O A 2010 *Phys. Rev. B* **81** 165105

[6] Trambly de Laissardière G, Mayou D and Magaud L 2012 *Phys. Rev. B* **86** 125413

[7] Trambly de Laissardière G, Mayou D and Magaud L 2010 *Nano Lett.* **10** 804

[8] Bistritzer R and MacDonald A H 2011 *Proc. Natl. Acad. Sci. U.S.A.* **108** 12233

[9] San-Jose P, Gonzalez J and Guinea F 2012 *Phys. Rev. Lett.* **108** 216802

[10] Suárez Morell E, Correa J D, Vargas P, Pacheco M and Barticevic Z 2010 *Phys. Rev. B* **82** 121407(R)

[11] Uchida K, Furuya S, Iwata J-I and Oshiyama A 2014 *Phys. Rev. B* **90** 155451

[12] Wang Z F, Liu F and Chou M Y 2012 *Nano Lett.* **12** 3833

[13] Weckbecker D, Shallcross S, Fleischmann M, Ray N, Sharma S and Pankratov O 2016 *Phys. Rev. B* **93** 035452

[14] Lopes dos Santos J M B, Peres N M R and Castro Neto A H 2012 *Phys. Rev. B* **86** 155449

[15] Sboychakov A O, Rakhmanov A L, Rozhkov A V and Nori F 2015 *Phys. Rev. B* **92** 075402

[16] Shallcross S, Sharma S and Pankratov O A 2008 *Phys. Rev. Lett.* **101** 056803

[17] Chu Z-D, He W-Y and He L 2013 *Phys. Rev. B* **87** 155419

[18] Bistritzer R and MacDonald A H 2010 *Phys. Rev. B* **81** 245412

[19] Moon P and Koshino M 2012 *Phys. Rev. B* **85** 195458

[20] Cherkez V, de Laissardière G T, Mallet P and Veuillen J Y 2015 *Phys. Rev. B* **91** 155428

[21] Luican A, Li G, Reina A, Kong J, Nair R R, Novoselov K S, Geim A K and Andrei E Y 2011 *Phys. Rev. Lett.* **106** 126802

[22] Yan W, Liu M, Dou R-F, Meng L, Feng L, Chu Z-D, Zhang Y, Liu Z, Nie J-C and He L 2012 *Phys. Rev. Lett.* **109** 126801

[23] Brihuega I, Mallet P, González-Herrero H, Trambly de Laissardière G, Ugeda M M, Magaud L, Gómez-Rodríguez J M, Ynduráin F and Veuillen J Y 2012 *Phys. Rev. Lett.* **109** 196802

[24] Li G, Luican A, Lopes dos Santos J M B, Castro Neto A H, Reina A, Kong J and Andrei E Y 2010 *Nat. Phys.* **6** 109

[25] Yin L-J, Qiao J-B, Zuo W-J, Li W-T and He L 2015 *Phys. Rev. B* **92** 081406(R)

[26] Yin L-J, Qiao J-B, Wang W-X, Zuo W-J, Yan W, Xu R, Dou R-F, Nie J-C and He L 2015 *Phys. Rev. B* **92** 201408(R)

[27] Wong D, Wang Y, Jung J, Pezzini S, DaSilva A M, Tsai H-Z, Jung H S, Khajeh R, Kim Y and Lee J 2015 *Phys. Rev. B* **92** 155409

[28] Yan W, Meng L, Liu M, Qiao J-B, Chu Z-D, Dou R-F, Liu Z, Nie J-C, Naugle D G and He L 2014 *Phys. Rev. B* **90** 115402

[29] Kim Y, Yun H, Nam S-G, Son M, Lee D S, Kim D C, Seo S, Choi H C, Lee H-J, Lee S W and Kim J S 2013 *Phys. Rev. Lett.* **110** 096602

[30] Miller D L, Kubista K D, Rutter G M, Ruan M, de Heer W A, First P N and Stroscio J A 2009 *Science* **324** 924

[31] Hass J, Varchon F, Millán-Otoya J E, Sprinkle M, Sharma N, de Heer W A, Berger C, First P N, Magaud L and Conrad E H 2008 *Phys. Rev. Lett.* **100** 125504

[32] Sprinkle M, Siegel D, Hu Y, Hicks J, Tejeda A, Taleb-Ibrahimi A, Le Fèvre P, Bertran F, Vizzini S, Enriquez H, Chiang S, Soukiassian P, Berger C, de Heer W A, Lanzara A and Conrad E H 2009 *Phys. Rev. Lett.* **103** 226803

[33] Koren E, Leven I, Lörtscher E, Knoll A, Hod O and Duerig U 2016 *Nat. Nanotech.* **11** 752

[34] Ni Z, Wang Y, Yu T, You Y and Shen Z 2008 *Phys. Rev. B* **77** 235403

[35] Angeli M, Mandelli D, Valli A, Amaricci A, Capone M, Tosatti E and Fabrizio M 2018 *Phys. Rev. B* **98** 235137

[36] Lin X and Tománek D 2018 *Phys. Rev. B* **98** 081410

[37] Carr S, Fang S, Po H C, Vishwanath A and Kaxiras E 2019 *Phys. Rev. Research* **1** 033072

[38] Goodwin Z A H, Corsetti F, Mostofi A A and Lischner J 2019 *Phys. Rev. B* **100** 235424

[39] Po H C, Zou L, Senthil T and Vishwanath A 2019 *Phys. Rev. B* **99** 195455

[40] Koshino M, Yuan N F Q, Koretsune T, Ochi M, Kuroki K and Fu L 2018 *Phys. Rev. X* **8** 031087

[41] Su Y and Lin S-Z 2018 *Phys. Rev. B* **98** 195101

[42] Tarnopolsky G, Kruchkov A J and Vishwanath A 2019 *Phys. Rev. Lett.* **122** 106405

[43] Julku A, Peltonen T J, Liang L, Heikkilä T T and Törmä P 2020 *Phys. Rev. B* **101** 060505

[44] Ledwith P J, Tarnopolsky G, Khalaf E and Vishwanath A 2020 *Phys. Rev. Research* **2** 023237

[45] Carr S, Fang S, Zhu Z and Kaxiras E 2019 *Phys. Rev. Research* **1** 013001

[46] Lucignano P, Alfè D, Cataudella V, Ninno D and Cantele G 2019 *Phys. Rev. B* **99** 195419

[47] Fang S, Carr S, Zhu Z, Massatt D and Kaxiras E 2019 *arXiv:1908.00058*

[48] Leconte N, Javvaji S, An J and Jung J 2019 *arXiv:1910.12805*

[49] Utama M I B, Koch R J, Lee K, Leconte N, Li H, Zhao S, Jiang L, Zhu J, Watanabe K, Taniguchi T, Ashby P D, Weber-Bargioni A, Zettl A, Jozwiak C, Jung J, Rotenberg E, Bostwick A and Wang F 2020 *Nat. Phys.*

[50] Lisi S, Lu X, Benschop T, de Jong T A, Stepanov P, Duran J R, Margot F, Cucchi I, Cappelli E and Hunter A 2020 *arXiv:2002.02289*

[51] Tomarken S L, Cao Y, Demir A, Watanabe K, Taniguchi T, Jarillo-Herrero P and

Ashoori R C 2019 *Phys. Rev. Lett.* **123** 046601

[52] Nuckolls K P, Oh M, Wong D, Lian B, Watanabe K, Taniguchi T, Bernevig B A and Yazdani A 2020 *arXiv:2007.03810*

[53] Zhang Z, Myers R, Watanabe K, Taniguchi T and LeRoy B J 2020 *arXiv:2003.09482*

[54] Wong D, Nuckolls K P, Oh M, Lian B, Xie Y, Jeon S, Watanabe K, Taniguchi T, Bernevig B A and Yazdani A 2020 *Nature* **582** 198

[55] Xie Y, Lian B, Jäck B, Liu X, Chiu C-L, Watanabe K, Taniguchi T, Bernevig B A and Yazdani A 2019 *Nature* **572** 101

[56] Choi Y, Kemmer J, Peng Y, Thomson A, Arora H, Polski R, Zhang Y, Ren H, Alicea J, Refael G, von Oppen F, Watanabe K, Taniguchi T and Nadj-Perge S 2019 *Nat. Phys.* **15** 1174

[57] Kerelsky A, McGilly L J, Kennes D M, Xian L, Yankowitz M, Chen S, Watanabe K, Taniguchi T, Hone J, Dean C, Rubio A and Pasupathy A N 2019 *Nature* **572** 95

[58] Jiang Y, Lai X, Watanabe K, Taniguchi T, Haule K, Mao J and Andrei E Y 2019 *Nature* **573** 91

[59] Ren Y-N, Lu C, Zhang Y, Li S-Y, Liu Y-W, Yan C, Guo Z-H, Liu C-C, Yang F and He L 2019 *arXiv:1912.07229*

[60] Li S-Y, Liu K-Q, Yin L-J, Wang W-X, Yan W, Yang X-Q, Yang J-K, Liu H, Jiang H and He L 2017 *Phys. Rev. B* **96** 155416

[61] Kim K, DaSilva A, Huang S, Fallahazad B, Larentis S, Taniguchi T, Watanabe K, LeRoy B J, MacDonald A H and Tutuc E 2017 *Proc. Natl. Acad. Sci. U.S.A.* **114** 3364

[62] Li S-Y, Zhang Y, Ren Y-N, Liu J, Dai X and He L 2019 *arXiv: 1912.13133*

[63] Polshyn H, Yankowitz M, Chen S, Zhang Y, Watanabe K, Taniguchi T, Dean C R and Young A F 2019 *Nat. Phys.* **15** 1011

[64] Serlin M, Tschirhart C L, Polshyn H, Zhang Y, Zhu J, Watanabe K, Taniguchi T, Balents L and Young A F 2020 *Science* **367** 900

[65] Cao Y, Fatemi V, Demir A, Fang S, Tomarken S L, Luo J Y, Sanchez-Yamagishi J D, Watanabe K, Taniguchi T, Kaxiras E, Ashoori R C and Jarillo-Herrero P 2018

*Nature* **556** 80

[66] Codecido E, Wang Q, Koester R, Che S, Tian H, Lv R, Tran S, Watanabe K, Taniguchi T, Zhang F, Bockrath M and Lau C N 2019 *Sci. Adv.* **5** eaaw9770

[67] Cao Y, Chowdhury D, Rodan-Legrain D, Rubies-Bigorda O, Watanabe K, Taniguchi T, Senthil T and Jarillo-Herrero P 2020 *Phys. Rev. Lett.* **124** 076801

[68] Cao Y, Fatemi V, Fang S, Watanabe K, Taniguchi T, Kaxiras E and Jarillo-Herrero P 2018 *Nature* **556** 43

[69] Arora H S, Polski R, Zhang Y, Thomson A, Choi Y, Kim H, Lin Z, Wilson I Z, Xu X, Chu J-H, Watanabe K, Taniguchi T, Alicea J and Nadj-Perge S 2020 *Nature* **583** 379

[70] Saito Y, Ge J, Watanabe K, Taniguchi T and Young A F 2020 *Nat. Phys.*

[71] Stepanov P, Das I, Lu X, Fahimniya A, Watanabe K, Taniguchi T, Koppens F H L, Lischner J, Levitov L and Efetov D K 2020 *Nature* **583** 375

[72] Lu X, Stepanov P, Yang W, Xie M, Aamir M A, Das I, Urgell C, Watanabe K, Taniguchi T, Zhang G, Bachtold A, MacDonald A H and Efetov D K 2019 *Nature* **574** 653

[73] Saito Y, Ge J, Rademaker L, Watanabe K, Taniguchi T, Abanin D A and Young A F 2020 *arXiv:2007.06115*

[74] Wu S, Zhang Z, Watanabe K, Taniguchi T and Andrei E Y 2020 *arXiv:2007.03735*

[75] Liu X, Wang Z, Watanabe K, Taniguchi T, Vafek O and Li J I A 2020 *arXiv:2003.11072*

[76] Cao Y, Rodan-Legrain D, Park J M, Yuan F N, Watanabe K, Taniguchi T, Fernandes R M, Fu L and Jarillo-Herrero P 2020 *arXiv:2004.04148*

[77] Uri A, Grover S, Cao Y, Crosse J A, Bagani K, Rodan-Legrain D, Myasoedov Y, Watanabe K, Taniguchi T, Moon P, Koshino M, Jarillo-Herrero P and Zeldov E 2020 *Nature* **581** 47

[78] Yankowitz M, Chen S, Polshyn H, Zhang Y, Watanabe K, Taniguchi T, Graf D, Young A F and Dean C R 2019 *Science* **363** 1059

[79] Zondiner U, Rozen A, Rodan-Legrain D, Cao Y, Queiroz R, Taniguchi T, Watanabe K, Oreg Y, von Oppen F, Stern A, Berg E, Jarillo-Herrero P and Ilani S 2020 *Nature*

**582** 203

[80] Das I, Lu X, Herzog-Arbeitman J, Song Z-D, Watanabe K, Taniguchi T, Bernevig B A and Efetov D K 2020 *arXiv:2007.13390*

[81] Sharpe A L, Fox E J, Barnard A W, Finney J, Watanabe K, Taniguchi T, Kastner M A and Goldhaber-Gordon D 2019 *Science* **365** 605

[82] Tschirhart C L, Serlin M, Polshyn H, Shragai A, Xia Z, Zhu J, Zhang Y, Watanabe K, Taniguchi T, Huber M E and F Y A 2020 *arXiv:2006.08053*

[83] Balents L, Dean C R, Efetov D K and Young A F 2020 *Nat. Phys.* **16** 725

[84] Liu Y-W, Qiao J-B, Yan C, Zhang Y, Li S-Y and He L 2019 *Phys. Rev. B* **99** 201408

[85] Zhang Y, Hou Z, Zhao Y-X, Guo Z-H, Liu Y-W, Li S-Y, Ren Y-N, Sun Q-F and He L 2020 *arXiv: 2002.10073*

[86] Wang W-X, Jiang H, Zhang Y, Li S-Y, Liu H, Li X, Wu X and He L 2017 *Phys. Rev. B* **96** 115434

[87] Koren E and Duerig U 2016 *Phys. Rev. B* **93** 201404(R)

[88] Ahn S J, Moon P, Kim T-H, Kim H-W, Shin H-C, Kim E H, Cha H W, Kahng S-J, Kim P, Koshino M, Son Y-W, Yang C-W and Ahn J R 2018 *Science* **361** 782

[89] Yao W, Wang E, Bao C, Zhang Y, Zhang K, Bao K, Chan C K, Chen C, Avila J, Asensio M C, Zhu J and Zhou S 2018 *Proc. Natl. Acad. Sci. U.S.A.* **115** 6928

[90] Moon P, Koshino M and Son Y-W 2019 *Phys. Rev. B* **99** 165430

[91] Park M J, Kim H S and Lee S 2019 *Phys. Rev. B* **99** 245401

[92] Yan C, Ma D-L, Qiao J-B, Zhong H-Y, Yang L, Li S-Y, Fu Z-Q, Zhang Y and He L 2019 *2D Mater.* **6** 045041

[93] Yu G, Wu Z, Zhan Z, Katsnelson M I and Yuan S 2019 *NPJ Comput. Mater.* **5** 1

[94] Deng B, Wang B, Li N, Li R, Wang Y, Tang J, Fu Q, Tian Z, Gao P, Xue J and Peng H 2020 *ACS Nano* **14** 1656

[95] Hua C-B, Chen R, Zhou B and Xu D-H 2020 *arXiv: 2001.07551*

[96] Yu G, Katsnelson M I and Yuan S 2020 *Phys. Rev. B* **102** 045113

[97] San-Jose P and Prada E 2013 *Phys. Rev. B* **88** 121408(R)

[98] Anđelković M, Covaci L and Peeters F M 2018 *Phys. Rev. Mater.* **2** 034004

[99] Efimkin D K and MacDonald A H 2018 *Phys. Rev. B* **98** 035404

[100] Huang S, Kim K, Efimkin D K, Lovorn T, Taniguchi T, Watanabe K, MacDonald A H, Tutuc E and LeRoy B J 2018 *Phys. Rev. Lett.* **121** 037702

[101] Rickhaus P, Wallbank J, Slizovskiy S, Pisoni R, Overweg H, Lee Y, Eich M, Liu M-H, Watanabe K, Taniguchi T, Ihn T and Ensslin K 2018 *Nano Lett.* **18** 6725

[102] Xu S G, Berdyugin A I, Kumaravadivel P, Guinea F, Krishna Kumar R, Bandurin D A, Morozov S V, Kuang W, Tsim B, Liu S, Edgar J H, Grigorieva I V, Fal'ko V I, Kim M and Geim A K 2019 *Nat. Commun.* **10** 4008

[103] De Beule C, Dominguez F and Recher P 2020 *arXiv:2005.05352*

[104] Qiao J-B, Yin L-J and He L 2018 *Phys. Rev. B* **98** 235402

[105] Liu Y-W, Su Y, Zhou X-F, Yin L-J, Yan C, Li S-Y, Yan W, Han S, Fu Z-Q, Zhang Y, Yang Q, Ren Y-N R and He L 2020 *arXiv:2007.01993*

[106] Ramires A and Lado J L 2018 *Phys. Rev. Lett.* **121** 146801

[107] Tsim B, Nam N N T and Koshino M 2020 *Phys. Rev. B* **101** 125409

[108] Walet N R and Guinea F 2019 *2D Mater.* **7** 015023

[109] Fleischmann M, Gupta R, Wullschläger F, Theil S, Weckbecker D, Meded V, Sharma S, Meyer B and Shallcross S 2020 *Nano Lett.* **20** 971

[110] De Beule C, Dominguez F and Recher P 2020 *arXiv:2003.08987*

[111] Hou T, Ren Y, Quan Y, Jung J, Ren W and Qiao Z 2020 *Phys. Rev. B* **101** 201403

[112] Chou Y-Z, Wu F and Sarma S D 2020 *arXiv:2004.15022*

[113] Sunku S S, Ni G X, Jiang B Y, Yoo H, Sternbach A, McLeod A S, Stauber T, Xiong L, Taniguchi T, Watanabe K, Kim P, Fogler M M and Basov D N 2018 *Science* **362** 1153

[114] Park M J, Kim Y, Cho G Y and Lee S 2019 *Phys. Rev. Lett.* **123** 216803

[115] Yoo H, Engelke R, Carr S, Fang S, Zhang K, Cazeaux P, Sung S H, Hovden R, Tsen A W and Taniguchi T 2019 *Nat. Mater.* **18** 448

[116] Wijk M M v, Schuring A, Katsnelson M I and Fasolino A 2015 *2D Mater.* **2** 034010

[117] Nam N N T and Koshino M 2017 *Phys. Rev. B* **96** 075311

[118] Gargiulo F and Yazyev O V 2017 *2D Mater.* **5** 015019

[119] Jain S K, Juričić V and Barkema G T 2016 *2D Mater.* **4** 015018

[120] McCann E 2006 *Phys. Rev. B* **74** 161403(R)

[121] Oostinga J B, Heersche H B, Liu X, Morpurgo A F and Vandersypen L M K 2008 *Nat. Mater.* **7** 151

[122] Li J, Wang K, McFaul K J, Zern Z, Ren Y, Watanabe K, Taniguchi T, Qiao Z and Zhu J 2016 *Nat. Nanotech.* **11** 1060

[123] Zhang Y, Tang T-T, Girit C, Hao Z, Martin M C, Zettl A, Crommie M F, Shen Y R and Wang F 2009 *Nature* **459** 820

[124] Castro E V, Novoselov K S, Morozov S V, Peres N M R, dos Santos J M B L, Nilsson J, Guinea F, Geim A K and Neto A H C 2007 *Phys. Rev. Lett.* **99** 216802

[125] Mak K F, Lui C H, Shan J and Heinz T F 2009 *Phys. Rev. Lett.* **102** 256405

[126] Vaezi A, Liang Y, Ngai D H, Yang L and Kim E-A 2013 *Phys. Rev. X* **3** 021018

[127] Zhang F, MacDonald A H and Mele E J J P o t N A o S 2013 *Proc. Natl. Acad. Sci.* **110** 10546

[128] Pelc M, Jaskólski W, Ayuela A and Chico L 2015 *Phys. Rev. B* **92** 085433

[129] Lee C, Kim G, Jung J and Min H 2016 *Phys. Rev. B* **94** 125438

[130] Alden J S, Tsen A W, Huang P Y, Hovden R, Brown L, Park J, Muller D A and McEuen P L 2013 *Proc. Natl. Acad. Sci.* **110** 11256

[131] Ju L, Shi Z, Nair N, Lv Y, Jin C, Velasco J, Ojeda-Aristizabal C, Bechtel H A, Martin M C, Zettl A, Analytis J and Wang F 2015 *Nature* **520** 650

[132] Yin L-J, Jiang H, Qiao J-B and He L 2016 *Nat. Commun.* **7** 11760

[133] Suzuki T, Iimori T, Ahn S J, Zhao Y, Watanabe M, Xu J, Fujisawa M, Kanai T, Ishii N, Itatani J, Suwa K, Fukidome H, Tanaka S, Ahn J R, Okazaki K, Shin S, Komori F and Matsuda I 2019 *ACS Nano* **13** 11981

[134] Lin Y-R, Samiseresht N, Franke M, Parhizkar S, Soubatch S, Amorim B, Lee T-L, Kumpf C, Tautz F and Bocquet F 2018 *arXiv:1809.07958*

[135] Takesaki Y, Kawahara K, Hibino H, Okada S, Tsuji M and Ago H 2016 *Chem. Mater.* **28** 4583

[136] Pezzini S, Mišeikis V, Piccinini G, Forti S, Pace S, Engelke R, Rossella F, Watanabe K, Taniguchi T, Kim P and Coletti C 2020 *Nano Lett.* **20** 3313

[137] Pal H K, Spitz S and Kindermann M 2019 *Phys. Rev. Lett.* **123** 186402

[138] Anderson P W 1987 *Science* **235** 1196

[139] Balents L 2010 *Nature* **464** 199

[140] Norman M R 2016 *Rev. Mod. Phys.* **88** 041002

[141] Fidrysiak M, Zegrodnik M and Spałek J 2018 *Phys. Rev. B* **98** 085436

[142] Guo H, Zhu X, Feng S and Scalettar R T 2018 *Phys. Rev. B* **97** 235453

[143] Guinea F and Walet N R 2018 *Proc. Natl. Acad. Sci. U.S.A.* **115** 13174

[144] Isobe H, Yuan N F Q and Fu L 2018 *Phys. Rev. X* **8** 041041

[145] Kennes D M, Lischner J and Karrasch C 2018 *Phys. Rev. B* **98** 241407

[146] Kang J and Vafek O 2018 *Phys. Rev. X* **8** 031088

[147] Liu C-C, Zhang L-D, Chen W-Q and Yang F 2018 *Phys. Rev. Lett.* **121** 217001

[148] Padhi B, Setty C and Phillips P W 2018 *Nano Lett.* **18** 6175

[149] Rademaker L and Mellado P 2018 *Phys. Rev. B* **98** 235158

[150] Venderbos J W F and Fernandes R M 2018 *Phys. Rev. B* **98** 245103

[151] Xu C and Balents L 2018 *Phys. Rev. Lett.* **121** 087001

[152] Yuan N F Q and Fu L 2018 *Phys. Rev. B* **98** 045103

[153] Alidoust M, Willatzen M and Jauho A-P 2019 *Phys. Rev. B* **99** 155413

[154] Chou Y-Z, Lin Y-P, Das Sarma S and Nandkishore R M 2019 *Phys. Rev. B* **100** 115128

[155] González J and Stauber T 2019 *Phys. Rev. Lett.* **122** 026801

[156] Huang T, Zhang L and Ma T 2019 *Sci. Bull.* **64** 310

[157] Kumar P, Vanhala T I and Törmä P 2019 *Phys. Rev. B* **100** 125141

[158] Lin Y-P and Nandkishore R M 2019 *Phys. Rev. B* **100** 085136

[159] Ray S, Jung J and Das T 2019 *Phys. Rev. B* **99** 134515

[160] Roy B and Juričić V 2019 *Phys. Rev. B* **99** 121407

[161] Tang Q-K, Yang L, Wang D, Zhang F-C and Wang Q-H 2019 *Phys. Rev. B* **99** 094521

[162] Wu F and Sarma S D 2019 *Phys. Rev. B* **99** 220507

[163] Wu X-C, Keselman A, Jian C-M, Pawlak K A and Xu C 2019 *Phys. Rev. B* **100** 024421

[164] You Y-Z and Vishwanath A 2019 *npj Quantum Materials* **4** 16

[165] Choi Y W and Choi H J 2018 *Phys. Rev. B* **98** 241412(R)

[166] Peltonen T J, Ojajärvi R and Heikkilä T T 2018 *Phys. Rev. B* **98** 220504

[167] Wu F, MacDonald A H and Martin I 2018 *Phys. Rev. Lett.* **121** 257001

[168] Lian B, Wang Z and Bernevig B A 2019 *Phys. Rev. Lett.* **122** 257002

[169] Wu F, Hwang E and Sarma S D 2019 *Phys. Rev. B* **99** 165112

[170] Angeli M, Tosatti E and Fabrizio M 2019 *Phys. Rev. X* **9** 041010

[171] Po H C, Zou L, Vishwanath A and Senthil T 2018 *Phys. Rev. X* **8** 031089

[172] Seo K, Kotov V N and Uchoa B 2019 *Phys. Rev. Lett.* **122** 246402

[173] Dodaro J F, Kivelson S A, Schattner Y, Sun X Q and Wang C 2018 *Phys. Rev. B* **98** 075154

[174] Ochi M, Koshino M and Kuroki K 2018 *Phys. Rev. B* **98** 081102(R)

[175] Liu S, Khalaf E, Lee J Y and Vishwanath A 2019 *arXiv:1905.07409*

[176] Xie M and MacDonald A H 2020 *Physical Review Letters* **124** 097601

[177] Alavirad Y and Sau J D 2019 *arXiv:1907.13633*

[178] Repellin C, Dong Z, Zhang Y-H and Senthil T 2020 *Phys. Rev. Lett.* **124** 187601

[179] Wu F and Sarma S D 2020 *Phys. Rev. Lett.* **124** 046403

[180] Zhang Y-H, Mao D and Senthil T 2019 *Phys. Rev. Research* **1** 033126

[181] Bultinck N, Chatterjee S and Zaletel M P 2020 *Phys. Rev. Lett.* **124** 166601

[182] Liu J and Dai X 2019 *arXiv:1911.03760*

[183] Chatterjee S, Bultinck N and Zaletel M P 2020 *Phys. Rev. B* **101** 165141

[184] Bultinck N, Khalaf E, Liu S, Chatterjee S, Vishwanath A and Zaletel M P 2019 *arXiv:1911.02045*

[185] Chichinadze D V, Classen L and Chubukov A V 2020 *arXiv:2007.00871*

[186] Zhang Y, Jiang K, Wang Z and Zhang F 2020 *Phys. Rev. B* **102** 035136

[187] Kang J and Vafek O 2019 *Phys. Rev. Lett.* **122** 246401

[188] Kang J and Vafek O 2020 *Phys. Rev. B* **102** 035161

[189] Da Liao Y, Meng Z Y and Xu X Y 2019 *Phys. Rev. Lett.* **123** 157601

[190] Sherkunov Y and Betouras J J 2018 *Phys. Rev. B* **98** 205151

[191] Cea T and Guinea F 2020 *Phys. Rev. B* **102** 045107

[192] Yao Q, van Bremen R, Slotman G J, Zhang L, Haartsen S, Sotthewes K,

Bampoulis P, de Boeij P L, van Houselt A, Yuan S and Zandvliet H J W 2017 *Phys. Rev. B* **95** 245116

[193] Cao Y, Luo J Y, Fatemi V, Fang S, Sanchez-Yamagishi J D, Watanabe K, Taniguchi T, Kaxiras E and Jarillo-Herrero P 2016 *Phys. Rev. Lett.* **117** 116804

[194] Goerbig M O 2011 *Rev. Mod. Phys.* **83** 1193

[195] Nomura K and MacDonald A H 2006 *Phys. Rev. Lett.* **96** 256602

[196] Yang K, Das Sarma S and MacDonald A H 2006 *Phys. Rev. B* **74** 075423

[197] Sheng L, Sheng D N, Haldane F D M and Balents L 2007 *Phys. Rev. Lett.* **99** 196802

[198] Nomura K, Ryu S and Lee D-H 2009 *Phys. Rev. Lett.* **103** 216801

[199] Lian Y, Rosch A and Goerbig M O 2016 *Phys. Rev. Lett.* **117** 056806

[200] Alicea J and Fisher M P A 2006 *Phys. Rev. B* **74** 075422

[201] Song Y J, Otte A F, Kuk Y, Hu Y, Torrance D B, First P N, De Heer W A, Min H, Adam S and Stiles M D 2010 *Nature* **467** 185

[202] Li S-Y, Zhang Y, Yin L-J and He L 2019 *Phys. Rev. B* **100** 085437

[203] Zhang Y, Jiang Z, Small J, Purewal M, Tan Y-W, Fazlollahi M, Chudow J, Jaszczak J, Stormer H and Kim P 2006 *Phys. Rev. Lett.* **96** 136806

[204] Jiang Z, Zhang Y, Stormer H L and Kim P 2007 *Phys. Rev. Lett.* **99** 106802

[205] Bolotin K I, Ghahari F, Shulman M D, Stormer H L and Kim P 2009 *Nature* **462** 196

[206] Du X, Skachko I, Duerr F, Luican A and Andrei E Y 2009 *Nature* **462** 192

[207] Dean C R, Young A F, Cadden-Zimansky P, Wang L, Ren H, Watanabe K, Taniguchi T, Kim P, Hone J and Shepard K L 2011 *Nat. Phys.* **7** 693

[208] Feldman B E, Krauss B, Smet J H and Yacoby A 2012 *Science* **337** 1196

[209] Young A F, Dean C R, Wang L, Ren H, Cadden-Zimansky P, Watanabe K, Taniguchi T, Hone J, Shepard K L and Kim P 2012 *Nat. Phys.* **8** 550

[210] Yu G L, Jalil R, Belle B, Mayorov A S, Blake P, Schedin F, Morozov S V, Ponomarenko L A, Chiappini F, Wiedmann S, Zeitler U, Katsnelson M I, Geim A K, Novoselov K S and Elias D C 2013 *Proc. Natl. Acad. Sci.* **110** 3282

[211] Bi Z, Yuan N F Q and Fu L 2019 *Phys. Rev. B* **100** 035448

[212] Wilson J H, Fu Y, Das Sarma S and Pixley J H 2020 *Phys. Rev. Research* **2** 023325

[213] He W-Y, Goldhaber-Gordon D and Law K T 2020 *Nat. Commun.* **11** 1650

[214] Zhu J, Su J-J and MacDonald A H 2020 *arXiv:2001.05084*

[215] Liu J and Dai X 2020 *NPJ Comput. Mater.* **6** 57

[216] Kim M, Xu S G, Berdyugin A I, Principi A, Slizovskiy S, Xin N, Kumaravadivel P, Kuang W, Hamer M, Krishna Kumar R, Gorbachev R V, Watanabe K, Taniguchi T, Grigorieva I V, Fal'ko V I, Polini M and Geim A K 2020 *Nature Communications* **11** 2339

[217] Wang Q-Y, Li Z, Zhang W-H, et al. 2012 *Chinese Phys. Lett.* **29** 037402

[218] Natterer F D, Ha J, Baek H, Zhang D, Cullen W G, Zhitenev N B, Kuk Y and Stroscio J A 2016 *Phys. Rev. B* **93** 045406

[219] Tonnoir C, Kimouche A, Coraux J, Magaud L, Delsol B, Gilles B and Chapelier C 2013 *Phys. Rev. Lett.* **111** 246805

[220] Qiao J-B, Gong Y, Zuo W-J, Wei Y-C, Ma D-L, Yang H, Yang N, Qiao K-Y, Shi J-A, Gu L and He L 2017 *Phys. Rev. B* **95** 201403

[221] Mayorov A S, Gorbachev R V, Morozov S V, Britnell L, Jalil R, Ponomarenko L A, Blake P, Novoselov K S, Watanabe K, Taniguchi T and Geim A K 2011 *Nano Lett.* **11** 2396

[222] Liu X, Hao Z, Khalaf E, Lee J Y, Ronen Y, Yoo H, Haei Najafabadi D, Watanabe K, Taniguchi T, Vishwanath A and Kim P 2020 *Nature* **583** 221

[223] Lee P A, Nagaosa N and Wen X-G 2006 *Rev. Mod. Phys.* **78** 17

[224] Keimer B, Kivelson S A, Norman M R, Uchida S and Zaanen J 2015 *Nature* **518** 179

[225] Scalapino D J 2012 *Rev. Mod. Phys.* **84** 1383

[226] Gurvitch M and Fiory A T 1987 *Phys. Rev. Lett.* **59** 1337

[227] Takagi H, Batlogg B, Kao H L, Kwo J, Cava R J, Krajewski J J and Peck W F 1992 *Phys. Rev. Lett.* **69** 2975

[228] Cooper R A, Wang Y, Vignolle B, Lipscombe O J, Hayden S M, Tanabe Y, Adachi T, Koike Y, Nohara M, Takagi H, Proust C and Hussey N E 2009 *Science* **323** 603

[229] Jin K, Butch N P, Kirshenbaum K, Paglione J and Greene R L 2011 *Nature* **476**

[230] Greene R L, Mandal P R, Poniatowski N R and Sarkar T 2020 *Annu. Rev. Condens. Matter Phys.* **11** 213

[231] González J and Stauber T 2020 *Phys. Rev. Lett.* **124** 186801

[232] Lawler M J, Fujita K, Lee J, Schmidt A R, Kohsaka Y, Kim C K, Eisaki H, Uchida S, Davis J C, Sethna J P and Kim E-A 2010 *Nature* **466** 347

[233] Daou R, Chang J, LeBoeuf D, Cyr-Choinière O, Laliberté F, Doiron-Leyraud N, Ramshaw B J, Liang R, Bonn D A, Hardy W N and Taillefer L 2010 *Nature* **463** 519

[234] da Silva Neto E H, Aynajian P, Frano A, Comin R, Schierle E, Weschke E, Gyenis A, Wen J, Schneeloch J, Xu Z, Ono S, Gu G, Le Tacon M and Yazdani A 2014 *Science* **343** 393

[235] Sato Y, Kasahara S, Murayama H, Kasahara Y, Moon E G, Nishizaki T, Loew T, Porras J, Keimer B, Shibauchi T and Matsuda Y 2017 *Nat. Phys.* **13** 1074

[236] Hanaguri T, Lupien C, Kohsaka Y, Lee D H, Azuma M, Takano M, Takagi H and Davis J C 2004 *Nature* **430** 1001

[237] Kohsaka Y, Taylor C, Fujita K, Schmidt A, Lupien C, Hanaguri T, Azuma M, Takano M, Eisaki H, Takagi H, Uchida S and Davis J C 2007 *Science* **315** 1380

[238] Fradkin E, Kivelson S A, Lawler M J, Eisenstein J P and Mackenzie A P 2010 *Annu. Rev. Condens. Matter Phys.* **1** 153

[239] Fernandes R M and Venderbos J W F 2019 *Sci. Adv.* **6** eaba8834

[240] Chichinadze D V, Classen L and Chubukov A V 2020 *Phys. Rev. B* **101** 224513

[241] Kozii V, Isobe H, Venderbos J W F and Fu L 2019 *Phys. Rev. B* **99** 144507

[242] Yang S L, Sobota J A, Howard C A, Pickard C J, Hashimoto M, Lu D H, Mo S K, Kirchmann P S and Shen Z X 2014 *Nat. Commun.* **5** 3493

[243] Ichinokura S, Sugawara K, Takayama A, Takahashi T and Hasegawa S 2016 *ACS Nano* **10** 2761

[244] Chapman J, Su Y, Howard C A, Kundys D, Grigorenko A N, Guinea F, Geim A K, Grigorieva I V and Nair R R 2016 *Sci. Rep.* **6** 23254

[245] Weller T E, Ellerby M, Saxena S S, Smith R P and Skipper N T 2005 *Nat. Phys.* **1** 39

[246] Margine E R, Lambert H and Giustino F 2016 *Sci. Rep.* **6** 21414

[247] Wu F and Das Sarma S 2019 *Phys. Rev. B* **99** 220507(R)

[248] Wu F 2019 *Phys. Rev. B* **99** 195114

[249] Chang C-Z, Zhang J, Feng X, et al. 2013 *Science* **340** 167

[250] Kou X, Guo S-T, Fan Y, Pan L, Lang M, Jiang Y, Shao Q, Nie T, Murata K, Tang J, Wang Y, He L, Lee T-K, Lee W-L and Wang K L 2014 *Phys. Rev. Lett.* **113** 137201

[251] Chang C-Z, Zhao W, Kim D Y, Zhang H, Assaf B A, Heiman D, Zhang S-C, Liu C, Chan M H W and Moodera J S 2015 *Nat. Mater.* **14** 473

[252] Lee I, Kim C K, Lee J, Billinge S J L, Zhong R, Schneeloch J A, Liu T, Valla T, Tranquada J M, Gu G and Davis J C S 2015 *Proc. Natl. Acad. Sci.* **112** 1316

[253] Mogi M, Yoshimi R, Tsukazaki A, Yasuda K, Kozuka Y, Takahashi K S, Kawasaki M and Tokura Y 2015 *Appl. Phys. Lett.* **107** 182401

[254] Yasuda K, Mogi M, Yoshimi R, Tsukazaki A, Takahashi K S, Kawasaki M, Kagawa F and Tokura Y 2017 *Science* **358** 1311

[255] Rozhkov A, Sboychakov A, Rakhmanov A and Nori F 2016 *Phys. Rep.* **648** 1

[256] Castro Neto A H, Guinea F, Peres N M R, Novoselov K S and Geim A K 2009 *Rev. Mod. Phys.* **81** 109

[257] McCann E and Koshino M 2013 *Rep. Prog. Phys.* **76** 056503

[258] Burg G W, Zhu J, Taniguchi T, Watanabe K, MacDonald A H and Tutuc E 2019 *Phys. Rev. Lett.* **123** 197702

[259] Cao Y, Rodan-Legrain D, Rubies-Bigorda O, Park J M, Watanabe K, Taniguchi T and Jarillo-Herrero P 2020 *Nature* **583** 215

[260] Shen C, Chu Y, Wu Q, Li N, Wang S, Zhao Y, Tang J, Liu J, Tian J, Watanabe K, Taniguchi T, Yang R, Meng Z Y, Shi D, Yazyev O V and Zhang G 2020 *Nat. Phys.* **16** 520

[261] He M, Li Y, Cai J, Liu Y, Watanabe K, Taniguchi T, Xu X and Yankowitz M 2020 *arXiv:2002.08904*

[262] Adak P C, Sinha S, Ghorai U, Sangani L D V, Watanabe K, Taniguchi T, Sensarma R and Deshmukh M M 2020 *Phys. Rev. B* **101** 125428

[263] Chebrolu N R, Chittari B L and Jung J 2019 *Phys. Rev. B* **99** 235417

[264] Choi Y W and Choi H J 2019 *Phys. Rev. B* **100** 201402(R)

[265] Lee J Y, Khalaf E, Liu S, Liu X, Hao Z, Kim P and Vishwanath A 2019 *Nat. Commun.* **10** 5333

[266] Wu F and Das Sarma S 2020 *Phys. Rev. B* **101** 155149

[267] Culchac F J, Del Grande R, Capaz R B, Chico L and Morell E S 2020 *Nanoscale* **12** 5014

[268] Samajdar R and Scheurer M S 2020 *Phys. Rev. B* **102** 064501

[269] Haddadi F, Wu Q, Kruchkov A J and Yazyev O V 2020 *Nano Lett.* **20** 2410

[270] Kerelsky A, Rubio-Verdú C, Xian L, Kennes D M, Halbertal D, Finney N, Song L, Turkel S, Wang L and Watanabe K 2019 *arXiv:1911.00007*

[271] Lee K, Utama M, Kahn S, Samudrala A, Leconte N, Yang B, Wang S, Watanabe K, Taniguchi T, Zhang G, Weber-Bargioni A, Crommie M, Ashby P D, Jung J, Wang F and Zettl A 2020 *arXiv:2006.04000*

[272] Halbertal D, Finney N R, Sunku S S, Kerelsky A, Rubio-Verdú C, Shabani S, Xian L, Carr S, Chen S and Zhang C 2020 *arXiv:2008.04835*

[273] Min H and MacDonald A H 2008 *Prog. Theor. Phys. Supp.* **176** 227

[274] Latil S and Henrard L 2006 *Phys. Rev. Lett.* **97** 036803

[275] Mak K F, Shan J and Heinz T F 2010 *Phys. Rev. Lett.* **104** 176404

[276] Yan J-A, Ruan W Y and Chou M Y 2011 *Phys. Rev. B* **83** 245418

[277] Wang W, Shi Y, Zakharov A A, Syväjärvi M, Yakimova R, Uhrberg R I G and Sun J 2018 *Nano Lett.* **18** 5862

[278] Yin L-J, Shi L-J, Li S-Y, Zhang Y, Guo Z-H and He L 2019 *Phys. Rev. Lett.* **122** 146802

[279] Bao W, Jing L, Velasco J, Lee Y, Liu G, Tran D, Standley B, Aykol M, Cronin S B, Smirnov D, Koshino M, McCann E, Bockrath M and Lau C N 2011 *Nat. Phys.* **7** 948

[280] Lui C H, Li Z, Mak K F, Cappelluti E and Heinz T F 2011 *Nat. Phys.* **7** 944

[281] Scherer M M, Uebelacker S, Scherer D D and Honerkamp C 2012 *Phys. Rev. B* **86** 155415

[282] Yin L-J, Wang W-X, Zhang Y, Ou Y-Y, Zhang H-T, Shen C-Y and He L 2017

*Phys. Rev. B* **95** 081402(R)

[283] Lee Y, Che S, Jr. J V, Tran D, Baima J, Mauri F, Calandra M, Bockrath M and Lau C N 2019 *arXiv:1911.04450*

[284] Chen G, Jiang L, Wu S, Lyu B, Li H, Chittari B L, Watanabe K, Taniguchi T, Shi Z, Jung J, Zhang Y and Wang F 2019 *Nat. Phys.* **15** 237

[285] Chen G, Sharpe A L, Fox E J, Zhang Y-H, Wang S, Jiang L, Lyu B, Li H, Watanabe K, Taniguchi T, Shi Z, Senthil T, Goldhaber-Gordon D, Zhang Y and Wang F 2020 *Nature* **579** 56

[286] Chittari B L, Chen G, Zhang Y, Wang F and Jung J 2019 *Phys. Rev. Lett.* **122** 016401

[287] Zhang Y-H and Senthil T 2019 *Phys. Rev. B* **99** 205150

[288] Pantaleon P A, Cea T, Brown R, Walet N R and Guinea F 2020 *arXiv:2003.05050*

[289] Scheurer M S and Samajdar R 2020 *Phys. Rev. Research* **2** 033062

[290] Zhang Y-H and Senthil T 2020 *arXiv:2003.13702*

[291] Qiao J-B and He L 2014 *Phys. Rev. B* **90** 075410

[292] Chen S, He M, Zhang Y-H, Hsieh V, Fei Z, Watanabe K, Taniguchi T, Cobden D H, Xu X and Dean C R 2020 *arXiv:2004.11340*

[293] Polshyn H, Zhu J, Kumar M A, Zhang Y, Yang F, Tschirhart C L, Serlin M, Watanabe K, Taniguchi T, MacDonald A H and Young A F 2020 *arXiv:2004.11353*

[294] Shi Y, Xu S, Ezzi M M A, Balakrishnan N, Garcia-Ruiz A, Tsim B, Mullan C, Barrier J, Xin N, Piot B A, Taniguchi T, Watanabe K, Carvalho A, Mishchenko A, Geim A K, Fal'ko V I, Adam S, Neto A H C and Novoselov K S 2020 *arXiv:2004.12414*

[295] Ma Z, Li S, Zheng Y-W, Xiao M-M, Jiang H, Gao J-H and Xie X C 2019 *arXiv:1905.00622*

[296] Park Y, Chittari B L and Jung J 2020 *arXiv:2005.01258*

[297] Rademaker L, Protopopov I and Abanin D 2020 *arXiv:2004.14964*

[298] Li X, Wu F and MacDonald A H 2019 *arXiv:1907.12338*

[299] Carr S, Li C, Zhu Z, Kaxiras E, Sachdev S and Kruchkov A 2020 *Nano Lett.* **20** 3030

[300] Mora C, Regnault N and Bernevig B A 2019 *Phys. Rev. Lett.* **123** 026402

[301] Tsai K-T, Zhang X, Zhu Z, Luo Y, Carr S, Luskin M, Kaxiras E and Wang K 2019 *arXiv:1912.03375*

[302] Zhu Z, Carr S, Massatt D, Luskin M and Kaxiras E 2020 *arXiv:2006.00399*

[303] Zuo W-J, Qiao J-B, Ma D-L, Yin L-J, Sun G, Zhang J-Y, Guan L-Y and He L 2018 *Phys. Rev. B* **97** 035440

[304] Liu J, Ma Z, Gao J and Dai X 2019 *Phys. Rev. X* **9** 031021

[305] Zhang Y-H, Mao D, Cao Y, Jarillo-Herrero P and Senthil T 2019 *Phys. Rev. B* **99** 075127

[306] Ma Z, Li S, Xiao M-M, Zheng Y-W, Lu M, Liu H, Gao J-H and Xie X 2020 *arXiv:2001.07995*

[307] Decker R, Wang Y, Brar V W, Regan W, Tsai H-Z, Wu Q, Gannett W, Zettl A and Crommie M F 2011 *Nano Lett.* **11** 2291

[308] Xue J, Sanchez-Yamagishi J, Bulmash D, Jacquod P, Deshpande A, Watanabe K, Taniguchi T, Jarillo-Herrero P and LeRoy B J 2011 *Nat. Mater.* **10** 282

[309] Yankowitz M, Xue J and LeRoy B J 2014 *J. Phys.: Condens. Matter* **26** 303201

[310] Dean C R, Young A F, Meric I, Lee C, Wang L, Sorgenfrei S, Watanabe K, Taniguchi T, Kim P, Shepard K L and Hone J 2010 *Nat. Nanotech.* **5** 722

[311] Yankowitz M, Ma Q, Jarillo-Herrero P and LeRoy B J 2019 *Nat. Rev. Phys.* **1** 112

[312] Kim Y, Balram A C, Taniguchi T, Watanabe K, Jain J K and Smet J H 2019 *Nat. Phys.* **15** 154

[313] Zhou H, Polshyn H, Taniguchi T, Watanabe K and Young A F 2020 *Nat. Phys.* **16** 154

[314] Yin J, Slizovskiy S, Cao Y, Hu S, Yang Y, Lobanova I, Piot B A, Son S-K, Ozdemir S, Taniguchi T, Watanabe K, Novoselov K S, Guinea F, Geim A K, Fal'ko V and Mishchenko A 2019 *Nat. Phys.* **15** 437

[315] Amet F, Bestwick A J, Williams J R, Balicas L, Watanabe K, Taniguchi T and Goldhaber-Gordon D 2015 *Nat. Commun.* **6** 5838

[316] Chiappini F, Wiedmann S, Novoselov K, Mishchenko A, Geim A K, Maan J C and Zeitler U 2015 *Phys. Rev. B* **92** 201412(R)

[317] Young A F, Sanchez-Yamagishi J D, Hunt B, Choi S H, Watanabe K, Taniguchi T, Ashoori R C and Jarillo-Herrero P 2014 *Nature* **505** 528

[318] Lee D S, Skákalová V, Weitz R T, von Klitzing K and Smet J H 2012 *Phys. Rev. Lett.* **109** 056602

[319] Yankowitz M, Xue J, Cormode D, Sanchez-Yamagishi J D, Watanabe K, Taniguchi T, Jarillo-Herrero P, Jacquod P and LeRoy B J 2012 *Nat. Phys.* **8** 382

[320] Gorbachev R V, Song J C W, Yu G L, Kretinin A V, Withers F, Cao Y, Mishchenko A, Grigorieva I V, Novoselov K S, Levitov L S and Geim A K 2014 *Science* **346** 448

[321] Wang E, Lu X, Ding S, Yao W, Yan M, Wan G, Deng K, Wang S, Chen G, Ma L, Jung J, Fedorov A V, Zhang Y, Zhang G and Zhou S 2016 *Nat. Phys.* **12** 1111

[322] Chen Z-G, Shi Z, Yang W, Lu X, Lai Y, Yan H, Wang F, Zhang G and Li Z 2014 *Nat. Commun.* **5** 4461

[323] Hunt B, Sanchez-Yamagishi J D, Young A F, Yankowitz M, LeRoy B J, Watanabe K, Taniguchi T, Moon P, Koshino M, Jarillo-Herrero P and Ashoori R C 2013 *Science* **340** 1427

[324] Ponomarenko L A, Gorbachev R V, Yu G L, Elias D C, Jalil R, Patel A A, Mishchenko A, Mayorov A S, Woods C R, Wallbank J R, Mucha-Kruczynski M, Piot B A, Potemski M, Grigorieva I V, Novoselov K S, Guinea F, Fal'ko V I and Geim A K 2013 *Nature* **497** 594

[325] Dean C R, Wang L, Maher P, Forsythe C, Ghahari F, Gao Y, Katoch J, Ishigami M, Moon P, Koshino M, Taniguchi T, Watanabe K, Shepard K L, Hone J and Kim P 2013 *Nature* **497** 598

[326] Ni G X, Wang H, Wu J S, Fei Z, Goldflam M D, Keilmann F, Özyilmaz B, Castro Neto A H, Xie X M, Fogler M M and Basov D N 2015 *Nat. Mater.* **14** 1217

[327] Dai S, Ma Q, Liu M K, Andersen T, Fei Z, Goldflam M D, Wagner M, Watanabe K, Taniguchi T, Thiemens M, Keilmann F, Janssen G C A M, Zhu S E, Jarillo-Herrero P, Fogler M M and Basov D N 2015 *Nat. Nanotech.* **10** 682

[328] Wang L, Zihlmann S, Liu M-H, Makk P, Watanabe K, Taniguchi T, Baumgartner A and Schönenberger C 2019 *Nano Lett.* **19** 2371

[329] Finney N R, Yankowitz M, Muraleetharan L, Watanabe K, Taniguchi T, Dean C

R and Hone J 2019 *Nat. Nanotech.* **14** 1029

[330] Kim K, Yankowitz M, Fallahazad B, Kang S, Movva H C P, Huang S, Larentis S, Corbet C M, Taniguchi T and Watanabe K 2016 *Nano Lett.* **16** 1989

[331] Ribeiro-Palau R, Zhang C, Watanabe K, Taniguchi T, Hone J and Dean C R 2018 *Science* **361** 690

[332] Yu L, Lee Y-H, Ling X, Santos E J G, Shin Y C, Lin Y, Dubey M, Kaxiras E, Kong J, Wang H and Palacios T 2014 *Nano Lett.* **14** 3055

[333] Massicotte M, Schmidt P, Vialla F, Schädler K G, Reserbat-Plantey A, Watanabe K, Taniguchi T, Tielrooij K J and Koppens F H L 2016 *Nat. Nanotech.* **11** 42

[334] Roy K, Padmanabhan M, Goswami S, Sai T P, Ramalingam G, Raghavan S and Ghosh A 2013 *Nat. Nanotech.* **8** 826

[335] Kośmider K, González J W and Fernández-Rossier J 2013 *Phys. Rev. B* **88** 245436

[336] Liu G-B, Shan W-Y, Yao Y, Yao W and Xiao D 2013 *Phys. Rev. B* **88** 085433

[337] Yuan H, Bahramy M S, Morimoto K, Wu S, Nomura K, Yang B-J, Shimotani H, Suzuki R, Toh M, Kloc C, Xu X, Arita R, Nagaosa N and Iwasa Y 2013 *Nat. Phys.* **9** 563

[338] Zeng H, Liu G-B, Dai J, Yan Y, Zhu B, He R, Xie L, Xu S, Chen X, Yao W and Cui X 2013 *Sci. Rep.* **3** 1608

[339] Zhang Y, Chang T-R, Zhou B, Cui Y-T, Yan H, Liu Z, Schmitt F, Lee J, Moore R, Chen Y, Lin H, Jeng H-T, Mo S-K, Hussain Z, Bansil A and Shen Z-X 2014 *Nat. Nanotech.* **9** 111

[340] Zhu Z Y, Cheng Y C and Schwingenschlögl U 2011 *Phys. Rev. B* **84** 153402

[341] Yang B, Tu M-F, Kim J, Wu Y, Wang H, Alicea J, Wu R, Bockrath M and Shi J 2016 *2D Mater.* **3** 031012

[342] Avsar A, Tan J Y, Taychatanapat T, Balakrishnan J, Koon G K W, Yeo Y, Lahiri J, Carvalho A, Rodin A S, O'Farrell E C T, Eda G, Castro Neto A H and Özyilmaz B 2014 *Nat. Commun.* **5** 4875

[343] Yang B, Lohmann M, Barroso D, Liao I, Lin Z, Liu Y, Bartels L, Watanabe K, Taniguchi T and Shi J 2017 *Phys. Rev. B* **96** 041409

[344] Antonio Benítez L, Sierra J F, Savero Torres W, Arrighi A, Bonell F, Costache M

V and Valenzuela S O 2018 *Nat. Phys.* **14** 303

[345] Ghiasi T S, Ingla-Aynés J, Kaverzin A A and van Wees B J 2017 *Nano Lett.* **17** 7528

[346] Omar S, Madhushankar B N and van Wees B J 2019 *Phys. Rev. B* **100** 155415

[347] Wang Z, Ki D-K, Khoo J Y, Mauro D, Berger H, Levitov L S and Morpurgo A F 2016 *Phys. Rev. X* **6** 041020

[348] Zihlmann S, Cummings A W, Garcia J H, Kedves M, Watanabe K, Taniguchi T, Schönenberger C and Makk P 2018 *Phys. Rev. B* **97** 075434

[349] Wakamura T, Reale F, Palczynski P, Guéron S, Mattevi C and Bouchiat H 2018 *Phys. Rev. Lett.* **120** 106802

[350] Island J O, Cui X, Lewandowski C, Khoo J Y, Spanton E M, Zhou H, Rhodes D, Hone J C, Taniguchi T, Watanabe K, Levitov L S, Zaletel M P and Young A F 2019 *Nature* **571** 85

[351] Zaletel M P and Khoo J Y 2019 *arXiv:1901.01294*

[352] Xi X, Wang Z, Zhao W, Park J-H, Law K T, Berger H, Forró L, Shan J and Mak K F 2016 *Nat. Phys.* **12** 139

[353] Guillamón I, Suderow H, Vieira S, Cario L, Diener P and Rodière P 2008 *Phys. Rev. Lett.* **101** 166407

[354] Galvis J A, Rodière P, Guillamon I, Osorio M R, Rodrigo J G, Cario L, Navarro-Moratalla E, Coronado E, Vieira S and Suderow H 2013 *Phys. Rev. B* **87** 094502

[355] Sipos B, Kusmartseva A F, Akrap A, Berger H, Forró L and Tutiš E 2008 *Nat. Mater.* **7** 960

[356] Lu J M, Zheliuk O, Leermakers I, Yuan N F Q, Zeitler U, Law K T and Ye J T 2015 *Science* **350** 1353

[357] de la Barrera S C, Sinko M R, Gopalan D P, Sivadas N, Seyler K L, Watanabe K, Taniguchi T, Tsen A W, Xu X, Xiao D and Hunt B M 2018 *Nat. Commun.* **9** 1427

[358] Gani Y S, Steinberg H and Rossi E 2019 *Phys. Rev. B* **99** 235404

[359] Li J, Leng H-B, Fu H, Watanabe K, Taniguchi T, Liu X, Liu C-X and Zhu J 2020 *Phys. Rev. B* **101** 195405

[360] Jiang Y, Anđelković M, Milovanović S P, Covaci L, Lai X, Cao Y, Watanabe K,

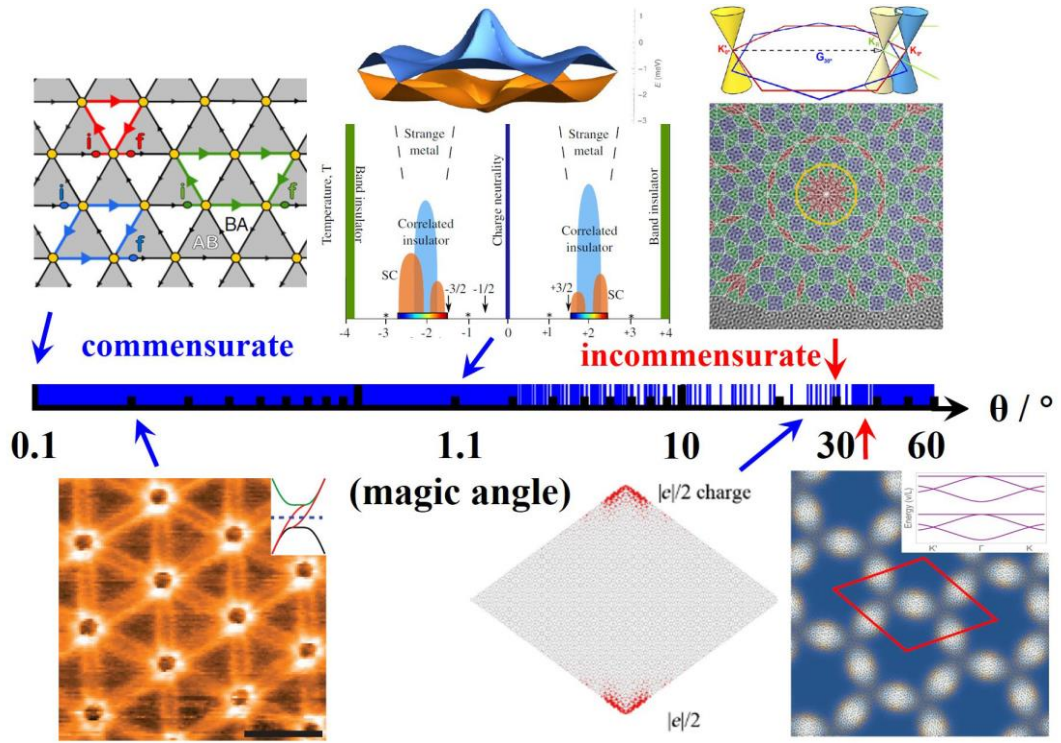

**Fig. 1 Abundant physical phenomena in TBG.** On the logarithmic axis of twist angle, the blue regions correspond to the commensurate angle, and the red regions correspond to the incommensurate angle. The arrow points out several typical physical phenomena, the angles from small to large correspond to: Aharonov–Bohm oscillations along the triangular network of AB/BA domain walls ($\theta \sim 0.1°$) [102], topologically protected helical edge states on the domain wall network ($\theta \sim 0.245°$) [100], strongly correlated phases for the partially filled moiré flat minibands on MATBG systems ($\theta \sim 1.1°$) [67, 143], the higher-order topological insulator with topological corner states ($\theta \sim 21.78°$) [114], the emergence of mirrored Dirac cones in graphene quasicrystal ($\theta \sim 30°$) [88, 89], flat bands caused by the electrons confined in a geometrically frustrated network of topologically protected modes ($\theta \sim 38.21°$) [137].

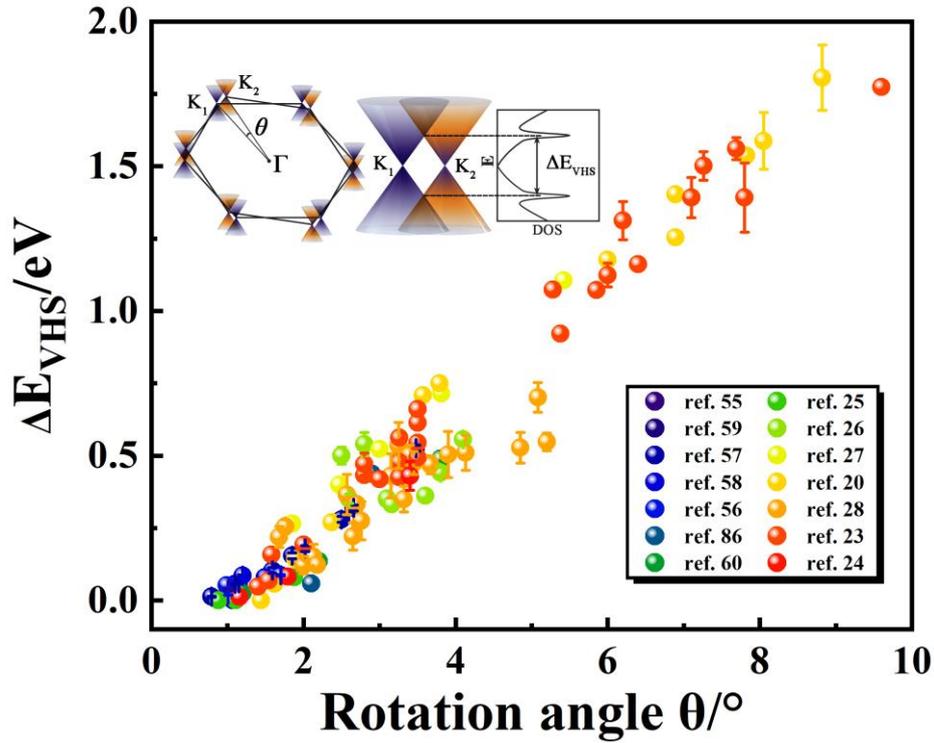

**Fig. 2 VHSs separation as a function of twist angle $\theta$.** Solid colorful circles are the experimental data measured in different TBG from different references [20, 23-28, 55-60, 86]. Error bars in energy represent the peak position change caused by slight doping change. The inset shows the electronic band structure of TBG. Left: The first Brillouin zone of TBG with twist angle $\theta$. $K_1$ and $K_2$ are the Dirac points of top and bottom layers. Middle: Energy dispersion relation of the overlap of the two Dirac cones, giving rise to two VHSs which generate peaks in the DOS. Right: Diagram of the energy dependence of DOS near the Fermi level.

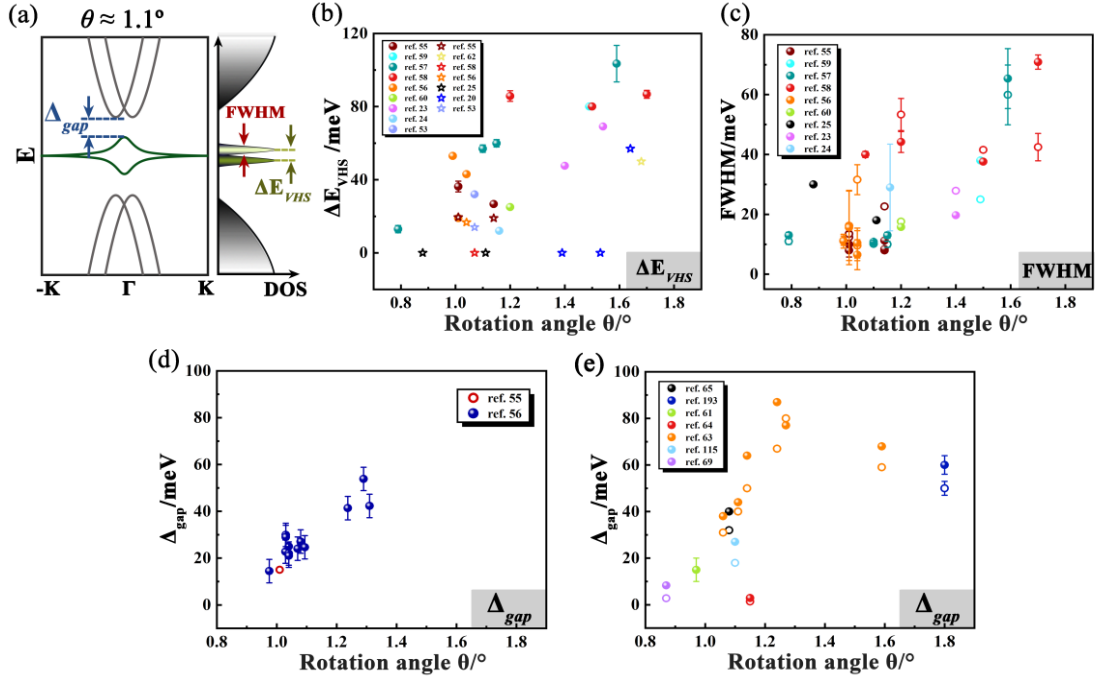

**Fig. 3 The sample differences of TBG in different references. (a)** A schematic diagram of the MATBG band structure and the corresponding DOS showing VHS peaks. The colored arrows point out three typical parameters: the band gap between the flat and the upper dispersive band ($\Delta_{gap}$ marked by blue), the full-width at half-maximum of the VHSs (FWHM marked by red), the energy separations of two VHSs ($\Delta E_{VHS}$ marked by green). **(b)** VHSs separation $\Delta E_{VHS}$ as a function of twist angle $\theta$ near MA-TBG. Solid colorful circles [23, 24, 53, 55-60] are the experimental data measured in different TBG, and the chemical potential is in between the two VHSs. The open stars symbols [20, 25, 53, 55, 56, 58, 62] represent data for the band becomes fully filled or fully unfilled. **(c)** The FWHM of the VHSs as a function of twist angle $\theta$, each of VHS is fully occupied or unoccupied. [23-25, 55-60]. Solid colorful circles represent the FWHM of conduction VHS. Open colorful circles represent FWHM of valence VHS. **(d)** and **(e)** The energy gap $\Delta_{gap}$ between the flat bands and the higher-energy dispersive bands as a function of twist angle $\theta$. (d) represents data for STM experiments [55, 56]. (e) represents data for transport experiments [61, 63-65, 69, 115, 193]. Solid colorful circles represent the gap between the electron flat band and the upper dispersive bands. Open colorful circles represent the gap between the hole flat band and the lower dispersive bands. Error bars in (b-e) originate from the slight doping change.

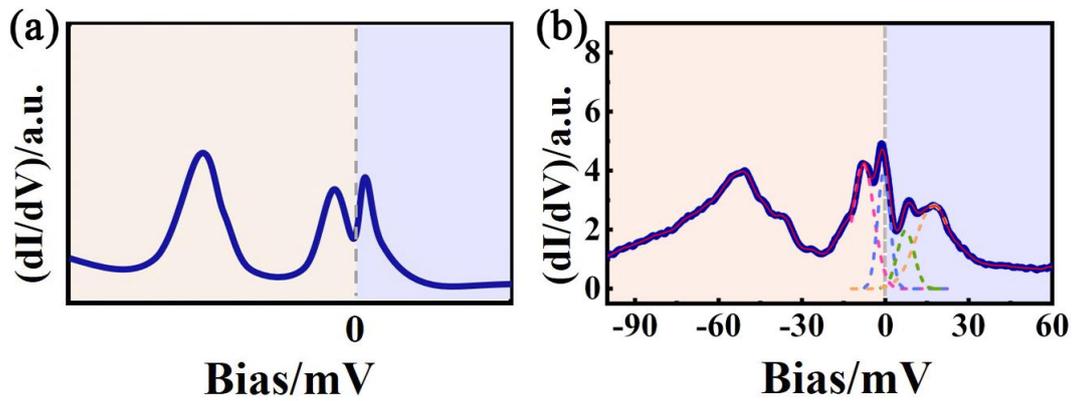

**Fig. 4 The STS spectra of TBG when one of the flat bands is partially filled. (a)** The schematic diagram of that the partially filled DOS peak in MATBG splits into two peaks. **(b)** The partially filled flat band splits into four DOS peaks in a non-magic-angle TBG with θ ~ 1.49º [59].

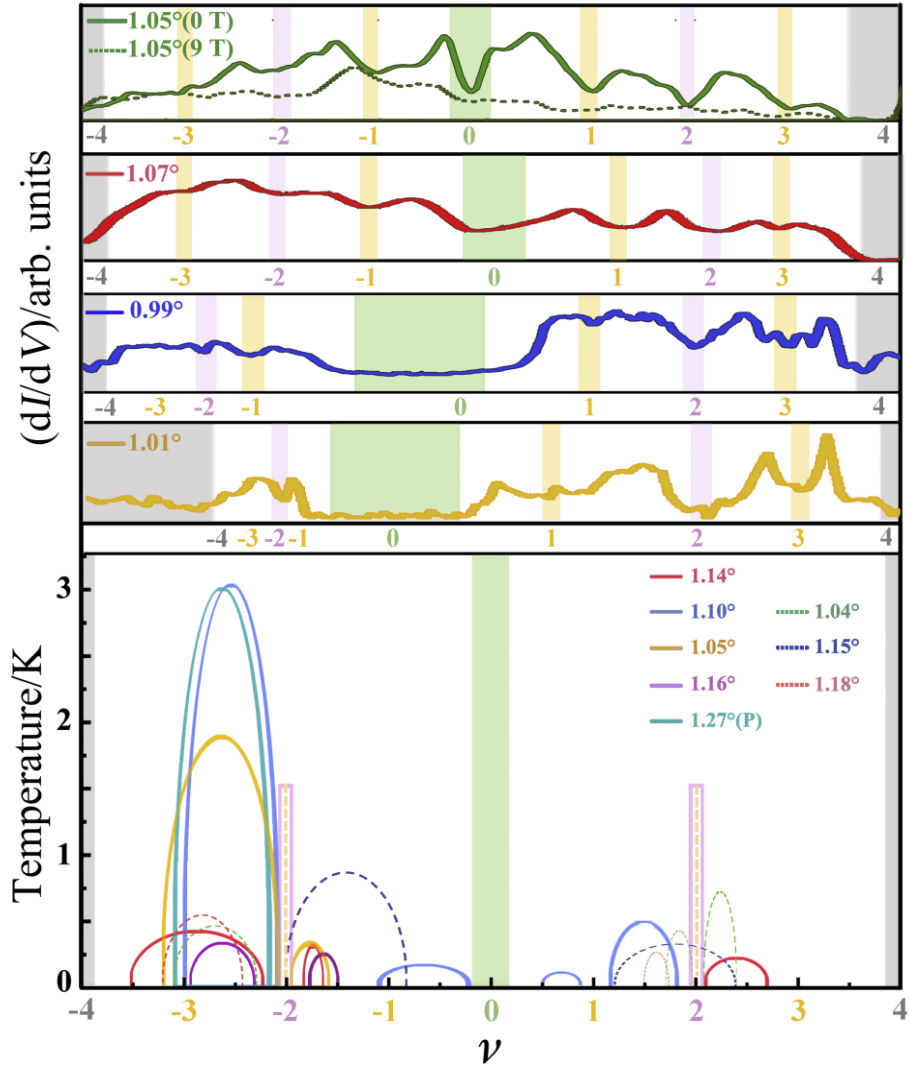

**Fig. 5 Correlated insulator and superconductivity in the MATBG. Top panel:** Conductance at the Fermi level as a function of filling factors $v$ in STM experiments. Grey areas correspond to fully occupied or unoccupied flat bands. Green areas correspond to the CNP ($v = 0$). Twist angles include 1.05° [54], 1.07° [58], 0.99° [56] and 1.01° [55]. **Bottom panel:** Superconducting phase diagrams acquired from the transport measurements. Solid superconducting domes indicate the coexistence of superconductivity and insulating states at $v = \pm 2$. Twist angles include 1.14° [78], 1.10° [72], 1.05° [68], 1.16° [68], 1.27° [78]. Dotted superconducting domes indicate that the superconductivity persists in the absence of the correlated insulators at $v = \pm 2$. Twist angles include 1.04° [71], 1.15° [71] and 1.18° [70].

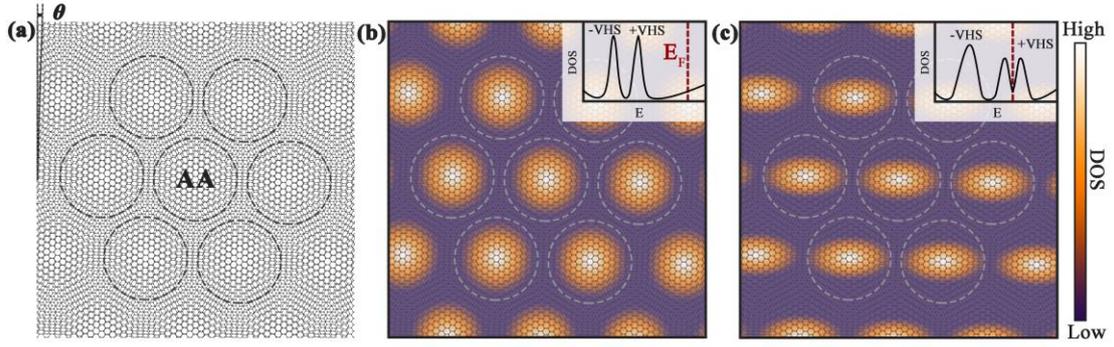

**Fig. 6 Spatially resolved conductance maps at the energies of the flat bands (VHSs) in MATBG.** **(a)** Schematic atomic structure of the TBG with a twist angle θ. The AA stacking configurations are marked in the panel, exhibiting the $C_6$ rotational symmetry. **(b)** The conductance maps at the energies around the VHS peaks show the same features as topographies and maintain the $C_6$ symmetry when the two VHSs are fully filled or empty. **(c)** The conductance maps at the energies around the VHS peaks show a pronounced anisotropy in each moiré when the Fermi level lies in one of the VHSs, reducing the symmetry from the initial $C_6$ to $C_2$.

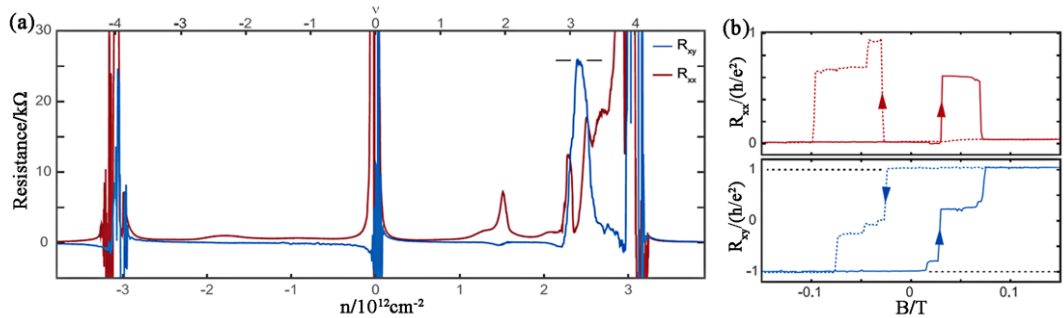

**Fig. 7 Quantized anomalous Hall effect in TBG (θ ~ 1.15º) at at ν =3. (a)** Longitudinal resistance $R_{xx}$ and Hall resistance $R_{xy}$ measured at $B$ =150 mT and $T$ = 1.6 K as a function of carrier density over the entire flat bands. Near the ν =3, $R_{xy}$ approaches $h/e^2$ and $R_{xx}$ reaches a deep minimum, illustrating the quantized anomalous Hall state. **(b)** Magnetic-field-dependent $R_{xx}$ and $R_{xy}$ at ν =3. The sweep directions are indicated by the arrows, showing an obvious hysteresis [64].

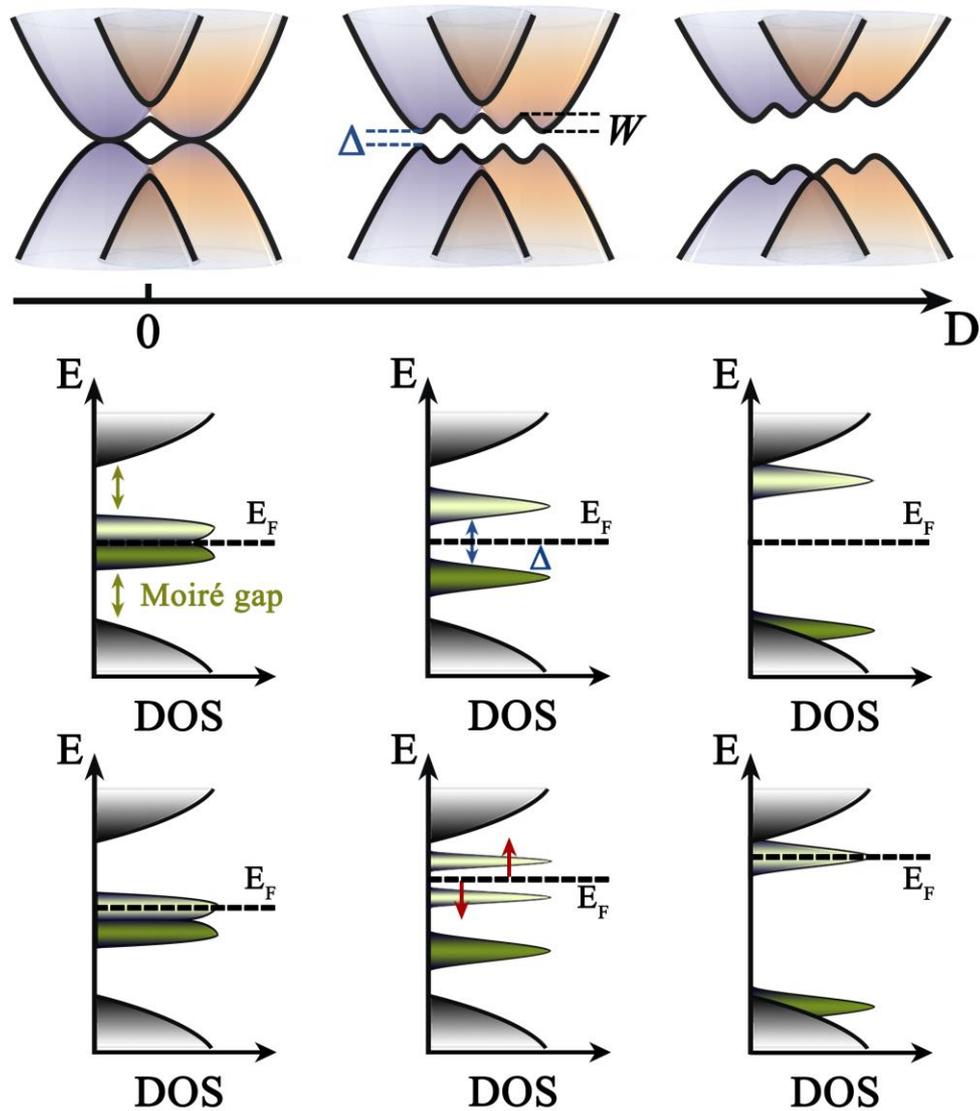

**Fig. 8 Schematic of the electrical-displacement-field-dependent low energy moiré bands and DOS in the TDBG.** The two parabolic dispersive bands are originated from the Bernal-stacked BLG. The two low-energy flat bands are separated from the high-energy dispersing bands by the moiré gaps. The bandwidth of the flat bands W and the band gap $\Delta$ at the charge neutral point can be easily tuned via the perpendicular electrical displacement field D. In a specific range of D, the correlated insulating states emerge when the isolated flat band is half filled, exhibiting the spin-polarized ordering.